\documentstyle[11pt,aaspp4]{article}
\def\msun {M$_{\odot}\ $}
\def\mic {$\mu$m\ }
\begin{document}
\title{THE EVOLUTION OF THE ELEMENTAL ABUNDANCES IN THE GAS AND DUST PHASES OF THE GALAXY}
\author{Eli Dwek \altaffilmark{1}}
\altaffiltext{1} {Laboratory for Astronomy and Solar Physics, Code 685, NASA/Goddard
Space Flight Center, Greenbelt, MD 20771.\ \ eli.dwek@gsfc.nasa.gov}

\begin{abstract}

We present models for the evolution
of the elemental abundances in the gas and dust phases of the interstellar medium (ISM) of our Galaxy by generalizing standard
models for its dynamical and chemical evolution. In these models, the stellar birthrate history is determined by
the infall rate of primordial gas, and by its functional dependence on the mass surface density of the stars and gas. We adopt a
two component model for the Galaxy, consisting of a central bulge and an exponential disk with different infall rates and
stellar birthrate histories. Condensation in stellar winds, Type Ia and Type II supernovae, and the accretion of refractory
elements onto preexisting grains in dense molecular clouds are the dominant contributors to the abundance of elements locked up
in the dust. Grain destruction by sputtering and evaporative grain-grain collisions in supernova remnants
are the most important mechanisms that return these elements back to the gas phase.

Guided by observations of dust formation in various stellar sources, and by the presence of isotopic anomalies in meteorites,
we calculate the production yield of silicate and carbon dust as a function of stellar mass. We find that Type II supernovae
are the main source of silicate dust in the Galaxy. Carbon dust is produced primarily by low mass stars in the
$\sim$ 2 - 5 \msun mass range. Type Ia SNe can be important sources of metallic iron dust in the ISM.

We also analyze the origin of the elemental depletion pattern, and find that the observed core + mantle depletion must reflect
 the efficiency of the accretion process in the ISM. We also find that grain destruction is very efficient, leaving only $\sim$
10\% of the refractory elements in grain cores. Observed core depletions are significantly higher, requiring significant UV,
cosmic ray, or shock processing of the accreted mantle into refractory core material.

Adopting the current grain destruction lifetimes from Jones et al. (1996), we formulate a prescription for its evolution in
time. We make a major assumption, that the accretion timescale evolves in a similar fashion, so that the current ratio
between these quantities is preserved over time. We then calculate the evolution of the dust abundance and composition at each
Galactocentric radius as a function of time.

We find that the dust mass is linearly proportional to the ISM metallicity, and is equal to about 40\% of the total mass of
heavy elements in the Galaxy, independent of Galactocentric radius. The derived relation of dust mass with metallicity 
is compared to the observed Galactic dust abundance gradient, and to the M$_{dust}$ versus log(O/H) relation that is observed
in external Dwarf galaxies. 

The dependence of dust
composition on the mass of the progenitor star, and the delayed recycling of newly synthesized dust by low mass stars back to
the ISM give rise to variations in the dust composition as a function of time. We identify three distinct epochs in the
evolution of the dust composition, characterized by different carbon-to-silicate mass ratios. Two such epochs are represented
by the Galaxy and the SMC. The third is characterized by an excess of carbon dust (compared to the Milky Way galaxy), and should
be observed in galaxies or star forming regions in which the most massive carbon stars are just evolving off the main sequence.

Our models
provide a framework for the self-consistent inclusion of dust in population synthesis models for various
pre-galactic and galactic systems, allowing for the calculation of their UV to far-infrared spectral energy
distribution at various stages of their evolution.

\end{abstract}

%**********************************************
\section{INTRODUCTION}
%**********************************************
Dust particles 
are formed in a wide variety of astrophysical environments, ranging from quiescent stellar outflows, including
various type giant and Wolf$-$Rayet stars, to explosive ejecta, including those of novae and Type Ia and Type II
supernovae. The presence of dust in these objects and in their surrounding media affects their spectral
appearance, and influences the determination of their underlying physical properties.

After their formation, dust particles are injected into the ISM where  they are subjected to a
variety of processes that affect their abundance and physical properties. Dust grains are subjected to 
 destructive processes including thermal and kinetic sputtering, evaporation due to thermal spiking, and
evaporative grain-grain collisions. Dust particles can also grow in the ISM by the accretion of ambient gas particles
and by accumulating into larger aggregates by coagulation.

The combined
effects of dust formation and processing in the ISM is manifested in the observed
elemental depletion pattern, which provides an important constraint on dust evolution models.
 Plotting the elemental depletions as a function of their condensation temperature, Field 
(1974) argued that the observed correlation 
reflects the dust formation efficiency in the various sources. The assumption behind this model is that any
further processing in the ISM will conserve the depletion pattern that was established in the sources. Snow (1975)
and Tabak (1979)
 showed that an equally good correlation exists between the elemental depletions and their
first ionization potential, arguing that the depletion pattern reflects the accretion
efficiency
 of the various elements in clouds. In this model all memory of the dust formation efficiency in the sources is
erased, and replaced by the elemental sticking efficiency inside
molecular clouds.

Accretion in clouds plays indeed an important role in determining the elemental depletion pattern. 
Plots depicting the depletions of select
 elements show that they are larger in the denser
phases of the ISM, and smaller in its warmer and less dense phases 
(see review by Savage \& Sembach 1996). This behavior is taken
as evidence for the  existence of two distinct dust "phases": (1) a core, consisting
of refractory material that is more immune to the various grain destruction processes
operating in the ISM; and (2) a mantle,
 consisting of a more loosely bound material that is accreted onto preexisting cores inside
dense molecular clouds. 

Another interpretation of the elemental depletion pattern was suggested by Dwek and Scalo (1980), who pointed out that the
sputtering threshold of various elements correlates with condensation temperature. Consequently, the elemental
depletion pattern may reflect the efficiency at which various elements are sputtered from the grains in the ISM.
Evidence for grain destruction in the ISM can be found from the depletions of
various elements in high velocity clouds. The trend of decreasing depletions with increasing
cloud velocity has been interpreted as evidence for 
 grain destruction in shocks (Spitzer 1976, Cowie 1978; Shull 1978). Further evidence for grain destruction in 
the general ISM can be inferred from the IR spectrum of shock$-$heated dust in supernova remnants (SNRs; e.g.
 Dwek \& Arendt 1992). Other processes that can selectively remove elements from the grains include classical evaporation,
which may be facilitated by temperature fluctuations, and chemical explosions. A summary of the various
 depletion theories which includes further references can be found in De Boer, Jura, \& Shull (1987) and 
Savage \& Sembach (1996).

Another process that effects the properties of dust in the ISM is coagulation in dense clouds. 
First evidence for this process was found by Jura (1980), who concluded that the below-average visual opacity through the 
$\rho$ Oph cloud complex is a result of grain coagulation in the cloud.
Further evidence to that effect can be inferred from the systematic differences between the extinction curves in the diffuse
intercloud medium and the outer cloud regions of the ISM (see Mathis 1990 for a review and further references). These
differences have been interpreted as evidence that inside these clouds dust particles accumulate into larger aggregates.
However, it is not clear whether these dust aggregates preserve this morphology in the more diffuse ISM, where even low
velocity shocks  can easily shatter them into their smaller grain constituents (Dominik, Jones, \& Tielens 1995). In any case, 
the process of coagulation only alters the grain size distribution but not the mass of refractory elements locked
up in dust, and will therefore not be considered in this paper.

Additional information on the sources, composition, and evolution of dust in the ISM can be
inferred from the isotopic composition of interstellar dust particles found in the
meteorites (see Zinner 1996 for a recent review on the subject). The isotopic composition of these dust
particles does not reflect the solar composition pattern, which represents a composition averaged
over many
 stellar sources. Instead, their isotopic composition reflects that
 of their distinct stellar sources and nuclear processes that synthesized the various
elements in these sources (Clayton 1982). 
 
An important constraint on dust
 models is that the abundance of elements required to be locked up  in solids should not exceed that available
in the interstellar medium. Consequently, a self consistent dust
evolution model must be part of a more general model that follows the elemental abundances and their isotopic composition in the
ISM as well. Such dust evolution models should incorporate all the various processes mentioned above: the dust formation
efficiencies and composition
 in the various sources, and the processing and cycling of dust in and between the various phases
 of the ISM.

Studies of the combined
effects of dust formation and grain processing in the ISM were conducted by Draine \& Salpeter (1979a, b), 
Dwek \& Scalo (1979, 1980), McKee (1989), and more recently by Jones,
Tielens, Hollenbach, \& McKee (1994) and Jones, Tielens, \& Hollenbach (1996). 
 An important result of these studies is that
grain destruction is very efficient, and that the rate at which elements are being removed
from the grains in the ISM exceeds their condensation rate in the various sources.
Further references on papers covering many
aspects of dust formation, abundances, and evolution can be found in the volume
"Interstellar Dust" (Allamandola \& Tielens 1989, eds.), and in the reviews by Dorschner \& Henning (1995), 
Tielens (1990), Tielens \& Allamandola (1987), Greenberg \& Hage (1990), Clayton (1982), and Draine (1990).

Various authors have constructed models to investigate different aspects of the evolution of interstellar
 dust. In particular, Dwek \&
Scalo (1980) developed a one-zone model to follow the evolution of refractory interstellar grains in the solar
neighborhood, generalizing
 existing models for the chemical evolution of the elements (e.g. Tinsley 1980).
The emphasis of their work was to examine the dependence of the interstellar depletions on
the dust production efficiency in various  sources and on the grain destruction lifetimes
in the ISM. Their work did not include the  effects of accretion in
clouds on the elemental depletions, or the effect of the delayed recycling of stellar 
nucleosynthesis products back to the ISM. Their model was extended by Liffman
\& Clayton (1989) and Liffman (1990), to include a two-phase ISM in which grain growth by
accretion and grain fragmentation were taken into account. The primary goals of these
models was to follow the grain size distribution, and the separate evolutionary histories of the refractory grain 
cores and mantles in order to identify the distribution and location in
the dust of the various carriers of the isotopic anomalies in meteorites. In a more recent
paper, Timmes \& Clayton (1996) followed specifically the evolution of SiC dust particles
in the ISM, in order to study the frequency distribution of its associated isotopic
anomalies in meteorites. 

In all these studies, the current abundance, composition, or size distribution of dust were the primary goal
of the investigations. A different approach was taken by Wang (1991a) who emphasized the
evolution of the dust abundance as a function of time. His model was essentially identical
to the Dwek$-$Scalo model, using the instantaneous
recycling approximation to describe the stellar production rate of the various elements and
the dust. The evolution of the dust plays an an important role in determining the opacity
of galaxies during their lifetime, and has therefore important cosmological implications.
The diffuse extragalactic infrared (IR) background contains the cumulative emission of
redshifted starlight and starlight that has been absorbed and reradiated by dust over the
cosmological history of the universe (e.g Franceschini et al. 1996). It is therefore affected by 
dust evolution in galaxies.
Furthermore, the presence of dust at high redshift can lead to erroneous conclusions
regarding galaxy number counts and the evolution of quasars (Wang 1991b, Pei \& Fall 1993). 
In an interesting test of dust evolution models, Ute \& Ferrara (1997) correlated the metallicity of a 
sample of dwarf galaxies with their dust content. The galaxy sample contains objects with different metallicities, 
and may therefore be providing a look at the temporal history of the dust evolution in these objects.

The current paper extends several aspects of the studies described above. 
The model presented here consists of three distinct components: (1) {\it a
dynamical model for the evolution of the stellar and gaseous contents of the Galaxy}. The 
disk mass grows by primordial infall to attain
its present mass and morphology. Different infall rates are assumed for the bulge and disk
component of the Galaxy. The stellar birthrate history is determined by the infall rate and
by the stellar and ISM mass densities of the disk; (2) 
{\it a chemical evolution model for the elements}. The formalism used here is essentially identical 
to the one used by
Matteuci \& Greggio (1986), Matteuci \& Fran\c{c}ois (1989), and more recently by Timmes, Woosley,
\& Weaver (1995). It follows the evolution of the elemental abundances, taking into account the delayed recycling
of the elements back into the ISM; and (3) {\it a
model for the evolution of the dust}. The basic  equations describing the chemical evolution of the elements are
generalized to include that of the dust. The dust is assumed to be either carbon rich or silicate type dust.
Two processes affecting the evolution of the dust,
 but not that of the elements, are included in the model: grain destruction by expanding SNRs, and grain
growth by accretion in molecular clouds. The ISM is considered to be a one$-$phase medium. Grain destruction and accretion
rates are therefore effective rates that represent values averaged over the cycling time of the dust
between the various ISM phases. Grain destruction in clouds by evaporative grain-grain collisions have a
negligible effect on the dust evolution and are ignored in this study. We also only follow the {\it mass} of dust in the
ISM, and do not calculate the evolution of the grain size distribution, as has been recently attempted by O'Donnell \& Mathis
(1997). We feel that such details are not warranted at this time, considering current uncertainties regarding the nature of
interstellar dust particles (Dwek 1997). One aspect that is followed in detail in the current model is the delayed recycling
of the elements and the dust back into the ISM. This delay has important observational consequences, and is responsible for
changes in the mixture of dust composition over time.

The mathematical
formalism of the model is described in detail in
\S 2. The evolution of the Galactic disk and that of the elemental composition are presented in \S 3 and \S 4, respectively. 
\S 5 describes the formation of dust and calculates the dust production rate by various sources. Timescales for grain
destruction by SNRs, and their time evolution are presented in
\S 6. The section includes also an brief discussion of the starburst galaxy M82, for which we derive rough estimates of grain
destruction lifetimes. Grain lifetimes for growth by accretion, and their time evolution are presented in \S 7. Model results
are presented in \S 8, and a brief summary and discussion of the paper is presented in \S.

%*************************************************************************
\section{MATHEMATICAL FORMULATION}
%*************************************************************************

\subsection{The Chemical Evolution of the ISM}

We consider a multi-zone model for the chemical evolution of our Galaxy,
in which star formation, gas infall, and expanding SNRs play the dominant role in
determining the abundances of the various elements in the gas and solid phases
of the ISM. Unlike conventional chemical evolution models (e.g. Matteucci 1997), we distinguish between the
abundances of the elements in the gas and solid phases of the ISM. We therefore define $\sigma_{ISM}$($A$,r,t) as
the {\it total} (gas + dust) surface mass density of a stable element
$A$ that is located at time $t$ at Galactocentric radius $r$. The ISM mass surface density, $\sigma_{ISM}$(r,t),
of {\it all} elements $A$ is given by the sum:

\begin{equation}
\sigma_{ISM}(r,t) = {\sum_{\{A\}} \sigma_{ISM}(A,r,t)}
\end{equation}

The equations for the evolution of $\sigma_{ISM}$(r,t) and $\sigma_{ISM}$($A$,r,t) at a given Galactocentric radius
$r$ and time $t$ are given by:

\begin{eqnarray}
{d\sigma_{ISM}(r, t)\over dt} & = &-{\cal B}(r,t)  \nonumber \\
& + &  \ \int_{M_l}^{M_u} 
{\cal B}(r,t-\tau (M))\ \phi(M)\ \left[{ M_{ej}(M,Z)\over M_{av}}\right]\ dM \\
& + & \left({d\sigma(r,t)\over dt}\right)_{inf} \nonumber
\end{eqnarray}

\noindent
and

\begin{eqnarray}
{d\sigma_{ISM}(A,r,t)\over dt} & = & -\ Z_{ISM}(A,r,t)\ {\cal B}(r,t)  \nonumber \\ 
          & + &\  \int_{M_l}^{M_u}{\cal B}(r,t-\tau(M))\ \phi (M)\ \left[{M_{ej}(A,M,Z)\over M_{av}}\right]\ dM   
\\
          & + & \left({d\sigma_A(r,t)\over dt}\right)_{inf} \nonumber 
\end{eqnarray}

\noindent 
where ${\cal B}$(r,t) is the stellar birthrate in units of M$_{\odot}$ pc$^{-2}$ Gyr$^{-1}$, and
$\tau(M)$ is the lifetime of a star of mass $M$. The function $\phi$(M) is the stellar mass spectrum (SMS),
defined as the {\it number} of stars born per unit mass, and normalized to unity in the \{M$_l$,
M$_u$\} mass interval. M$_{ej}$(M,Z) and
M$_{ej}$($A$,M,Z) are, respectively, the total mass and the mass of element $A$, that are ejected by a star of
mass $M$ and initial metallicity $Z$ back into the ISM, M$_{av}$ is the SMS-averaged mass of the newly born stars, and 
(${d\sigma\over dt})_{inf}$, is the infall rate of primordial gas into the Galaxy. Finally, the parameter
Z$_{ISM}$($A$,r,t) in equation (3) is the mass  fraction of a given element in the ISM, defined as:

\begin{equation}
Z_{ISM}(A,r,t) \equiv {\sigma_{ISM}(A,r,t)\over \sigma_{ISM}(r,t)}
\end{equation}

\noindent
 Further details on
the choice of SMS and various related parameters are given in \S 3 below.

Following Matteucci \& Greggio (1986; see also
Matteucci \& Fran\c cois 1989; Chiappini, Matteucci, \& Gratton 1997; and Timmes, Woosley, \& Weaver 1995), we will
rewrite  equation (3) as an explicit sum of its various contributions:
\begin{eqnarray}
{d\sigma_{ISM}(A,r,t)\over dt} & = & -\ Z_{ISM}(A,r,t)\ {\cal B}(r,t)  \nonumber \\
          & + & \ \int_{M_l}^{M_{b1}}  {\cal B}(r,t-\tau (M)) \ \phi (M)\ \left[{M_{ej}(A,M,Z)\over M_{av}}\right]
                    \ dM \nonumber \\ 
          & + & \beta\ \left({M_{ej}^I(A)\over M_{av}}\right)\ \int_{M_{b1}}^{M_{b2}}\phi(M_b)\ dM_b
               \int_{\mu_m}^{1\over 2} f(\mu)\ {\cal B}(r,t-\tau(\mu M_b))\ d\mu  \\
					     & + & (1-\beta)\int_{M_{b1}}^{M_{b2}}{\cal B}(r,t-\tau(M))\ \phi(M)\ \left[{M_{ej}(A,M,Z)\over M_{av}}
		              \right]\ dM \nonumber \\
          & + & \int_{M_{b2}}^{M_u}{\cal B}(r,t-\tau(M))\ \phi(M)\ \left[{M_{ej}(A,M,Z)\over M_{av}}\right]\ dM
                    \nonumber \\
          & + & \left({d\sigma(A,r,t)\over dt}\right)_{inf} \nonumber 
\end{eqnarray}
\noindent
The first term in equation (5) represents the removal rate of element
$A$ from the ISM due to astration. The second term represents the enrichment rate of the element due to stars in
the
\{M$_l$, M$_{b1}$\} mass range. The third equation represents the enrichment rate of $A$ due to binary systems
that become Type Ia supernovae, where M$_{ej}^I$(A) is the mass of element $A$ synthesized in the explosion, and
M$_{b1}$ and M$_{b2}$ are, respectively, the lower and upper mass limits of binary systems that become Type Ia
SNe. The parameter
$\beta$ determines the relative rates of Type Ia to Type II events, and will be discussed in more detail below. The
fourth term represents the enrichment due to binary systems that do not undergo Type Ia events, and stars with
masses $\geq M_w$ that become Type II SNe. The fifth term represents the enrichment rate of element $A$ due to
massive stars (M $\geq$ M$_w$) that become Type II SNe, and the final term is the accretion rate of element $A$ onto the
disk due to infall. Throughout this work, we assume that the infalling gas has primordial abundances. We note that the
mathematical formalism is unchanged in the presence of {\it outflows}. The last term will then simply represent
the net rate at which the abundance of element $A$ is changing in the disk due to combined effect of accretion and
outflow.

\subsection{The Chemical Evolution of the Dust}

The evolution of an element $A$ in the dust can be described in a way similar to that of the elements in the
ISM (eq. 5) as:

\begin{eqnarray}
{d\sigma_{dust}(A,r,t)\over dt} & = & - Z_{dust}(A,r,t))\ {\cal B}(r,t)  \nonumber \\
          & + & \int_{M_l}^{M_{b1}}  {\cal B}(t-\tau (M))\ \phi (M)\ \left[{\delta_{cond}^w(A) M_{ej}(A,M,Z) \ 
                \over M_{av}}\right]\ dM \nonumber \\ 
          & + & \beta\ \left({\delta_{cond}^I(A) M_{ej}(A)\over M_{av}}\right)\ \int_{M_{b1}}^{M_{b2}}\phi(M_b)\
dM_b
               \int_{\mu_m}^{1\over 2} f(\mu){\cal B}(r,t-\tau(\mu M_b))\ d\mu  \nonumber \\
					     & + & (1-\beta)\int_{M_{b1}}^{M_{b2}}{\cal B}(r,t-\tau(M))\ \phi(M)\ \left[{\delta_{cond}^w(A)
	                 M_{ej}(A,M,Z)	\over M_{av}}\right]\ dM  \nonumber \\
          & + & \int_{M_{b2}}^{M_u}{\cal B}(r,t-\tau(M))\ \phi(M)\ \left[{\delta_{cond}^{II}(A) M_{ej}(A,M,Z)
                 \over M_{av}}\right]\ dM   \\ 
          & - & \sigma_{dust}(A,r,t)/\tau_{SNR}(A,r,t) \nonumber \\
          & + & \sigma_{dust}(A,r,t)\ \left(1-{\sigma_{dust}(A,r,t)\over \sigma_{ISM}(A,r,t)}\right)/\ \tau_{accr}(A,r,t)
                  \nonumber \\
          & - & \left({d\sigma_{dust}(A,r,t)\over dt}\right)_{outf} \nonumber  
\end{eqnarray}
\noindent
where 
\begin{equation}
Z_{dust}(A,r,t) \equiv {\sigma_{dust}(A,r,t)\over \sigma_{ISM}(r,t)} \nonumber
\end{equation}

\noindent
is the mass fraction of element $A$ locked up in dust, and the parameters
$\delta_{cond}^w(A)$,
$\delta_{cond}^I(A)$, and $\delta_{cond}^{II}(A)$ represent the condensation efficiency of the
element $A$ in stellar winds, Type I SNe, and Type II SNe, respectively. Eq (6) introduces two new terms that
affect the evolution of dust in the ISM: (1) the sixth term in the equation, which is the rate at which the element
$A$ is returned to the gas phase as the dust is destroyed by SNRs; and (2) the
seventh term, which represents the rate at which an element $A$ is removed from the gas phase by accretion onto
{\it preexisting} dust particles in molecular clouds. The parameters $\tau_{SNR}(A,r,t)$,
and $\tau_{accr}(A,r,t)$,
 are, respectively, the timescales for these processes, and will be discussed in more detail below.

%*********************************************
\section{THE EVOLUTION OF THE GALACTIC DISK}
%*********************************************
\subsection{The Infall Model}
We adopt a dynamical model for the Galaxy, which starts from a zero mass disk that grows by accretion of
primordial gas from its surrounding medium. The present (t=t$_G$) disk morphology consists of a central bulge, and
an exponential disk. The present mass distribution of the disk is assumed to be exponential of
the form:

\begin{equation}
\sigma_{disk}(r, t_G) = \sigma_{\odot}\exp\left(-\ {r-R_{\odot}\over R_{disk}}\right)\ \ \ \ \ \ \ \ \ 0 \leq
r(kpc)
\leq 15
\end{equation}

\noindent
where $\sigma_{\odot}$, is the current mass density of the disk at the solar
circle, and
$R_{disk}$ is the exponential scale length of the disk. Current estimates for $\sigma_{\odot}$ are in the
40 $-$ 80
M$_{\odot}$ pc$^{-2}$ range (Gould, Bahcall, \& Flynn 1996). Here we adopt a compromise value of
$\sigma_{\odot}$ = 60 M$_{\odot}$ pc$^{-2}$. The scale length of the disk is taken to be 3.5 kpc, which is a fit to
the Bahcall \& Soneira (1980) model, and was also used by Wainscoat et al. (1992) in their faint source model of the
Galaxy. The total mass of the disk, integrated out to a radius of 15 kpc is $M_{disk} \approx\ 2\pi\ R_{disk}^2\
\sigma_{\odot}\exp(R_{\odot}/R_{disk})$, where R$_{disk}$ is in kpc. For a value of $\sigma_{\odot}=\ 60\ M_{\odot}\
pc^{-2}$, the disk mass is 5.2 10$^{10}\ M_{\odot}$.

For the bulge we adopt a oblate spheroidal mass distribution of the form $\rho(x) = \rho_0
\exp(-x^2)$ (model G0 in Dwek et al. 1995), up to a galactic radius of 2 kpc, which is about equal to the
corotation radius beyond which stable orbits cannot exist (Binney et al. 1991). The disk-projected mass surface
density of this oblate spheroid can be written as:

\begin{eqnarray}
\sigma_{bulge}(r, t_G) & = & \sigma_0\ \exp\left(-{r^2\over R^2_{bulge}}\right)\ \ \ \ for\ \  r\ \leq 2\ kpc \nonumber
\\
                       & = & 0  \ \ \ \ \ \ \ \ \ \ \ \ \ \ \ \ \ \ \ \ \ \ \ for\ \ \ r \geq 2\ kpc
\end{eqnarray}

\noindent
where $\sigma_0$ is the central mass density of the bulge, and $R_{bulge}$ a scaling parameter. The total mass of
the bulge, integrated to a distance of 2 kpc, is $M_{bulge}=\pi R^2_{bulge}\ \sigma_0\
\left[1-\exp(-4/R^2_{bulge})\right]$, where R$_{bulge}$ is in kpc. From a fit of the sky-projected intensity
distribution of the oblate spheroid to the observed near-IR intensity of the bulge observed by the {\it COBE}/DIRBE,
Dwek et al. (1995) find a value of
$R_{bulge} \approx$ 1.3 kpc, and a total photometrically deduced bulge mass of $\approx 1.3\ 10^{10}$ \msun. This mass
is very similar to the dynamically deduced bulge mass of $\approx 1\ 10^{10}$ \msun (Kent 1992). Adopting a bulge mass
of
$1.3\ 10^{10}$ \msun, we get that the resulting central mass density for the bulge is $\sigma_0 = 2.7\ 10^3$ \msun
pc$^{-2}$.

The total surface mass density of the disk at the present epoch, $\sigma_{tot}(r,t_G)$ is:

\begin{equation}
\sigma_{tot}(r, t_G)= \sigma_{disk}(r, t_G) + \sigma_{bulge}(r, t_G)
\end{equation}

\noindent
giving a total Galactic mass of 6.5 10$^{10}\ M_{\odot}$. We note that unlike TWW95,
we did not impose any continuity between the values or slopes of the surface mass densities of the disk and bulge at the
co-rotation radius, since it has no physical basis.

The Galactic infall rate at each distance $r$ is assumed to be decreasing exponentially with time:

\begin{equation}
\left({d\sigma(r, t)\over dt}\right)_{inf} = A(r) \exp\left[-{t\over \tau_{inf}(r)}\right]
\end{equation}

\noindent
where $\tau_{inf}(r)$ is a radially dependent timescale of the form given by CMG96:
\begin{eqnarray}
\tau_{inf}(r) & = & 0.4615\ r(kpc) + 0.077\ \ \ Gyr \ \ \ \  for\ \ \ r \geq 2\ kpc \nonumber  \nl
              & = & 1\ \ Gyr \ \ \ \ \ \ \ \  for\ \ \ r \leq 2\ kpc 
\end{eqnarray}

\noindent
Equation (12) gives an infall rate of 1 Gyr for the formation of the bulge, and an infall rate of 6 Gyr in the solar
neighborhood.
       
\subsection{The Stellar Birthrate}
We will assume that the stellar birthrate ${\cal B}$(r,t), defined
here as the mass of stars born per unit disk area per unit time, can be written as the product:
\begin{eqnarray}
{\cal B}(r,t) & = & {\cal C}(r,t)\ \int_{M_l}^{M_{up}}\ M\ \phi (M)\ dM \nonumber \\
              & = & {\cal C}(r,t) M_{av}
\end{eqnarray}
\noindent
where $\phi (M)$, the stellar mass spectrum (SMS), is the number of stars born per unit mass and is normalized to unity
in the \{M$_l$, M$_{up}$\} mass interval,
${\cal C}(r,t)$ is a stellar creation function, and M$_{av}$ is the average mass of stars in the \{M$_l$, M$_{up}$\}
mass interval. Details of the SMS are described in \S 3.3 below. In particular, 
${\cal R}$(r,t, M$_1$,M$_2$), the {\it number} of stars born in the \{M$_1$, M$_2$\} mass interval per unit disk area
per unit time can be written in terms of ${\cal B}$(r,t) as:
\begin{eqnarray}
{\cal R}(r,t,M_1,M_2) & = &  {\cal C}(r,t)\ \int_{M_l}^{M_{up}}\ \phi (M)\ dM \nonumber \\
   & = & {{\cal B}(r,t)\over M_{av}}\ \int_{M_1}^{M_2}\ \phi (M)\ dM
\end{eqnarray}

Following Dopita \&
Ryder (1994) and Ryder (1995) we will write the stellar birthrate in the form ${\cal B}
\sim \sigma_{tot}^k\ \sigma_{ISM}^n$. In the classical Schmidt law: ${\cal
B} \sim \rho_{gas}^2$, where $\rho_{gas}$ represents the number density of the gas. The additional dependence
on the total mass density of the disk proposed by Dopita \& Ryder (1994) presumably includes the feedback
effects of the star-formation/death  processes on the birthrate. The birthrate, in units of
M$_{\odot}$ pc$^{-2}$ Gyr$^{-1}$, is then:

%------------
\begin{equation}
{\cal B}(r,t) = \nu\ \sigma_{tot}(r,t) \left({\sigma_{tot}(r,t)\over \sigma_{tot}(R_{\odot},t)}\right)^k
		                                  \left({\sigma_{ISM}(r,t)\over \sigma_{tot}(r,t_G)}\right)^n
\end{equation}
%-------------

\noindent
where $\nu$ is the star formation efficiency factor in Gyr$^{-1}$. At each galactic radius r, the ISM mass surface
density is normalized to the total mass surface density at the current epoch, t$_G$, and at each epoch $t$, the
total mass surface density is normalized to that in the solar vicinity.  The present birthrate at the solar circle
is thus given by:
%-------------
\begin{equation}
{\cal B}(R_{\odot},t_G) \equiv {\cal B}_{\odot} = \nu\ \sigma_{\odot}\ 
					                          \left({\sigma_{ISM}(R_{\odot},t_G)\over \sigma_{\odot}}\right)^n
\end{equation}
%-------------

\noindent
where the last term represents the present ISM mass fraction in the solar vicinity. The present birthrate
as a function of Galactic radius is given by:
%------------
\begin{equation}
{\cal B}(r,t_G) = \nu\ \sigma_{tot}(r,t_G) \left({\sigma_{tot}(r,t_G)\over
\sigma_{\odot}}\right)^k		                                 
\end{equation}
%-------------

\noindent
where as in equation (8), $\sigma_{\odot}\equiv \sigma_{tot}(R_{\odot},t_G)$. 

The observational constraints on ${\cal B}_{\odot}$,
$\sigma_{\odot}$, and the present day mass fraction of the ISM (see Table 1) constrain the allowable values of $\nu$, $k$, and
$n$. Following CMG96, we adopted the values of $\nu$ = 1.0 Gyr$^{-1}$, and \{k, n\} = \{0.5, 1.5\} in this paper.

\subsection{The Evolution of the Disk and the ISM}
A characteristic of the infall model is the gradual buildup of the disk from zero mass
to its present composition of stars and gas. The relative mix of stars and gas is determined by the exact
prescription for the stellar birthrate, and the flux of infalling material. Figure 1 presents the stellar
birthrate per unit surface area at the center of the Galactic bulge and in the solar circle, as a function of time. Also
 plotted in the figure is the disk averaged value of the stellar birthrate. Stellar
birthrates increase initially, as more material is accreted onto the Galaxy, decreasing at a later epoch as the infalling gas
is converted to stars. Since the infall timescale onto the bulge is only 1 Gyr, star formation peaks at the Galactic center at
an early epoch of about 0.6 Gyr. At the solar circle, where the infall timescale was taken to be 4 Gyr, the stellar birthrate
peaks at about 6 Gyr. Figure 2 depicts the radial dependence of the stellar birthrate as a function of Galactocentric radius.
Figure 3 shows the evolution of the surface density of the stars and the ISM at the Galactic center and the solar circle as a
function of time. Table 2
summarizes the various model output parameters at the solar circle. As shown in the table, the model reproduces the
observational constraints in the solar neighborhood.

%****************************************
\section{THE EVOLUTION OF THE ELEMENTS}
%****************************************

The evolution of the elements is determined by the stellar mass spectrum (SMS), and the stellar nucleosynthesis
yield.

\subsection{The Stellar Mass Spectrum (SMS)}
The SMS was originally parameterized as a single power law of the form
$\simÊM^{-\alpha}$ with $\alpha$ = 2.35 over the 0.5 to 10 \msun  mass interval (Salpeter 1955). 
Subsequent studies (Miller
\& Scalo 1979; Scalo 1986) showed that the Salpeter SMS significantly overestimates the number
of stars with masses below $\sim$ 1 \msun , and that its slope at higher masses is somewhat steeper than
the Salpeter one, and characterized by a value of $\alpha$ = 2.7 instead. The slope for M $>$ 1 \msun was
confirmed by subsequent studies (Rana \& Basu 1992) who derived a value of $\alpha$ = 2.6, however, Van
Buren (1985) derived a somewhat shallower slope with an $\alpha$ = 2.0. Recent studies of Tsujimoto et al. (1997)
suggest a slope of $\alpha$ between 2.3 and 2.6 above 1 \msun. The shape of the SMS is more uncertain at lower
masses. Scalo's results suggests the existence of a distinct maxima in the SMS around $\sim$ 0.2 \msun, decreasing
at lower masses, with
$\alpha
\approx$ 1.6. The results of Tinney (1993) suggest a flattening of the SMS in the $\sim$ 0.12 to 0.25 \msun  mass
range with a possible upturn at lower masses, whereas the studies of Kroupa, Tout, \& Gilmore (1993) suggest a shallower
power with an average value of $\alpha$ = 1.6 between 0.1 and 1.0 M$_{\odot}$. Fortunately, the results of our
calculations are not sensitive to the shape of the SMS below $\sim$ 1 M$_{\odot}$. The functional form finally adopted
for the SMS in the
\{M$_l$, M$_{u}$\} = \{ 0.1 \msun , 40 \msun \} mass range is:

\begin{eqnarray}
\phi(M) \   = &  C_1            &\ \   for \ \ \ \ \   	   0.1  \leq  M/M_{\odot}   \leq  0.3 \nonumber \\
		      =  & C_2  M^{-1.6}     &\ \  	for \ \ \ \ \    0.3  \leq  M/M_{\odot}   \leq   1  \\
		      =  & C_3  M^{-2.6}     &\ \  	for\ \ \ \ \ 	  1    \leq  M/M_{\odot}   \leq   40 \nonumber
\end{eqnarray}

\noindent
where the constants C$_1$, C$_2$, and C$_3$ are determined from the normalization:

\begin{equation}
\int_{M_1}^{M_{u}} \phi(M)\ dM = 1
\end{equation}
and by the requirement that $\phi$(M) be continuous at M = 0.3 and 1 M$_{\odot}$ .

\subsection{The Type I and II Supernovae Rate}

Type II SNe result from the collapse of the iron core of stars with masses above a 
value M$_w$, taken here to be equal to 8 M$_{\odot}$. The rate of SN II, in units of
pc$^{-2}$ Gyr$^{-1}$ is therefore given from equation (14) by:

\begin{equation}
{\cal R}_{SNII}(r,t) =  M_{av}^{-1}\ \int_{M_w}^{M_u}  {\cal B}(r,t-\tau(M))\ \phi(M)\ dM
\end{equation}

Type Ia supernovae are the result of thermonuclear explosions in
accreting white dwarfs in binary systems (e.g. Hachisu, Kato, \& Nomoto 1996, and references therein). Let 
M$_b = m+M$  be the mass of the binary system, where $m \leq M$, and let $\mu \equiv m/M_b$. The SN Ia rate,
${\cal R}_{SNI}$, given in pc$^{-2}$ Gyr$^{-1}$, can then be written as the following double integral (Greggio \&
Renzini 1983, Matteucci \& Greggio 1986):

%--------------
\begin{equation}
{\cal R}_{SNIa}(r,t) = \beta\ M_{av}^{-1}\ \int_3^{16} \phi(M_b)\ dM_b \int_{\mu_m}^{1/2} f(\mu)\
B(r,t-\tau(M=\mu M_b))\ d\mu
\end{equation}
%--------------
where  $\mu_m$ = max\{$M_{TO}(t)/M_b,\
(M_b-M_w)/M_b$\}, and M$_{TO}$ is the main-sequence turnoff mass at time $t$. The function f($\mu$)=24$\mu^2$ is
normalized to unity in the \{0, ${1\over 2}$\} interval, and represents the probability that a binary system will
have a mass ratio $\mu$. Equation (21) assumes a minimum and maximum binary mass of M$_{b1}$=3 M$_{\odot}$, and
M$_{b2}$ = 2 M$_w$ = 16 M$_{\odot}$, respectively. 

The parameter $\beta$ determines the value of R$_{Ia/II}\equiv {\cal R}_{SNIa}(R_{\odot},t_G)/{\cal
R}_{SNII}(R_{\odot},t_G)$, the currently observed  ratio of SN Ia to SN II events.
The value of R$_{Ia/II}$ can span a wide range of values, and Rocca-Volmerange \&
Schaeffer (1990) considered the effects of varying R$_{Ia/II}$ from 0.25 to 1.22 on Galactic chemical evolution. CMG96 found
that a value of R$_{Ia/II}$ = 0.37 best reproduces the elemental abundances, whereas Tsujimoto et al. (1995) and TWW95
derive  similar values of R$_{Ia/II}$ = 0.16 for the relative SN Ia and SN II rates. In our model, the value of
$\beta$ was adjusted until it produced a good fit to the solar abundances at the epoch the solar system was formed.

\subsection{Nucleosynthesis Yields}

Stellar nucleosynthesis yields are commonly divided into four categories: (1) yields from low-mass stars
(M$\leq$ M$_w$ = 8 M$_{\odot}$) that return mass quiescently in the form of stellar winds and planetary nebulae
back to the ISM; (2) yields from binary stars in the \{M$_{b1}$, M$_{b2}$\} mass range
that become Type Ia SNe; (3) yields from stars with masses $\geq$ M$_w$ that become Type II SNe; and (4) yields
from stars with masses $\gtrsim$ 40 \msun, whose evolution is dominated by stellar mass loss.

Yields for the stars in the 1 to 8 \msun mass range were taken from Renzini \& Voli (1981; hereafter RV81) from
their tabulated values for metallicities of Z = 0.004 and Z = Z$_{\odot}$ = 0.020, with $\alpha$ = 3/2 and $\eta$ = 1/3. 

For Type Ia SNe we used the yields for model W7, tabulated in Thielemann, Nomoto \& hashimoto (1993), which are based
on the model W7 of Nomoto, Thielemann \& Yokoi (1984), and Thielemann, Nomoto, \& Yokoi (1986).

For stars in the 11 to 40 \msun mass range we adopted the post-explosive nucleosynthesis of
Woosley \& Weaver (1995). In addition to the metallicity of the progenitor stars, the yields depend on the
explosion energy, and following Timmes, Woosley
\& Weaver (1995) we adopted models "A" for stars with initial masses below 25 M$_{\odot}$, and
models "B" for the 30, 35, and 40 M$_{\odot}$ stars.

For the primordial abundances of the infalling gas we adopted the Big Bang composition with Z(H) = 0.77, and
Z(He)=0.23 (Olive \& Steigman 1995).

\subsection{Observational Tests of the Chemical Evolution Model} 
Viable chemical evolution models should be able to reproduce the following observational constraints (e.g. Edvardsson et al.
(1993): (1) the observed solar abundances at the time the solar system formed; (2) the observed
age-metallicity relation; (3) the Galactic abundance gradient; (4) the frequency distribution of metallicity (also known
as the G-dwarf problem); and (5) the [$\alpha$/Fe] versus [Fe/H] abundance trend. Figure 4 tests our model against
these  various constraints, illustrating that the model reproduces them as well as other comparable models (e.g
CMG97 or TWW95). Figure 4a shows that the model reproduces the elemental abundances at the time the sun was formed, with the
exception of N. Nitrogen is overproduced as a result of the HBB process. The overproduction of N will not effect the results
of this paper, since N is neither condensed in stellar outflows, nor accreted onto dust in the ISM. Figure 4b compares the
model age$-$metallicity relation versus observational constraints, and Figure 4c compares the metallicity frequency predicted
by the model to that inferred from G-dwarf population. All these tests do not depend on the fraction of the metals in the ISM
is locked up in dust. However, the Galactic metal abundance gradient does, since the observations sample only the gas phase
abundance of the elements in the ISM. Figure 4d plots the total ISM [Fe/H] abundance, [Fe/H]$\equiv$(Fe/H)/(Fe/H)$_{\odot}$, as
well as the gas phase [Fe/H] abundance, versus Galactic radius. Also plotted in the figure are the observational constraints.
The figure clearly shows the improvement in the fit when the fraction of the iron locked up in the dust is subtracted from the
total Fe abundance.  The final observational test, the [$\alpha$/Fe] versus [Fe/H] abundance trend, has been discussed
extensively elsewhere (TWW95). Since we adopted similar stellar SMS and abundance yields, our trends should be similar to
theirs.
 Figure 5 shows the current SN Ia and SN II supernovae rates at these locations as a function of time. As shown in Table 2,
the current SN II and SN Ia rates are consistent with observational constraints.

%************************************
\section{THE EVOLUTION OF THE DUST}
%************************************
\subsection{Dust Formation}

Dust formation takes place in a variety of astronomical sources that can be broadly divided into two
categories: sources that undergo quiescent mass loss, and sources that return their ejecta explosively back to the
ISM. Below we present the various lines of evidence for the formation of dust in the various sources. The prescription
used to calculate the dust production yield in the various sources is given in \S3.8 and a comparison between the
calculated dust production/destruction rates at the current epoch and observational estimates is presented in Table
3.    

\subsubsection{Quiescent Formation: Low-Mass Stars, M $<$ 8 \msun} 

The cool atmospheres of late-type giant and supergiant stars provide an ideal environment for the nucleation
and growth of refractory grains (e.g. Sedlmayr 1989 and references therein). The presence of dust in these
objects is inferred from the presence of an excess of IR emission above the underlying stellar continuum, and from
the extinction, scattering, and polarization of the underlying stellar UV-visual photons.  The composition of the dust
particles around these stars is inferred from from their spectral features: the 9.7 and 18 \mic silicate features, the
11.3 \mic SiC feature, a broad $\sim$ 30 \mic MgS feature, the ice absorption feature at 3.1 $\mu$m, and a featureless
spectrum usually associated with graphite of amorphous carbon dust. The type of dust that forms in the outflowing
matter depends on the C/O  ratio in the ejecta. When C/O $>$ 1, all the oxygen is tied up in CO molecules, and the
newly formed grains will be carbon rich. If conversely, C/O
$<$ 1, the excess oxygen will combine with other refractory elements to form silicate type material (e.g. Whittet
1989, and references therein). There is observational evidence supported by laboratory data that MgS and sulfites form
when C/O $\approx$ 1 (Nuth et al. 1985), but these dust species are relatively rare, and will not be
considered here.

\subsubsection{Quiescent Formation: Wolf-Rayet Stars}

High mass stars (M $\gtrsim$ 40 \msun) can lose mass at a rate of $\sim (2-10)\ 10^{-5}$ \msun yr$^{-1}$ (Abbott \&
Conti 1987; van der Hucht 1992), which will have a significant effect on their structure and evolution (e.g. Maeder
1991). Above a critical mass loss rate these stars will strip off their H-rich envelope, exposing the underlying
cores. Such stripped cores will appear as Wolf-Rayet stars of type N if the exposed material is N-rich, and of
type WC, if the exposed material is C-rich. However, although rapid mass loss is a common characteristic of all WR
stars, only the coolest WC stars (WC8 and WC9) exhibit an IR excess that can be attributed to the formation of
dust in their atmosphere (Cohen, Barlow, \& Kuhi 1975; Williams, van der Hucht, \& Th\'e 1987). The dust spectrum
is featureless, suggesting carbon type dust, consistent with the enrichment of carbon in the stellar ejecta. Since
these stars are rare, their contribution to the overall production of interstellar dust is not expected to be
significant. WR originate from stars with masses in excess of $\sim$ 40 \msun, and are therefore by a factor
of $\sim$ 20  less frequent than Type II SNe. For a local Type II rate of 0.024 pc$^{-2}$ Gyr$^{-1}$ (see Table 2),
 we get a WR rate of $\sim 2\ 10^{-3}$ pc$^{-2}$ Gyr$^{-1}$. Adopting a total wind mass of $\sim$ 10 \msun, and
 a carbon dust-to-mass ratio of $\sim$ 0.02, the total carbon dust production rate from WR stars is $\sim 2\ 10^{-4}$
M$_{\odot}$ pc$^{-2}$ Gyr$^{-1}$. 
   
\subsubsection{Following Explosive Conditions: Novae}

Many novae eruptions exhibit a spectacular evolution of their lightcurve characterized by the
development of a steep decline at UV-visual wavelengths, about 3 to 4 months after the outburst, accompanied
by a rapid rise at infrared wavelengths (see Gehrz 1988 for an extensive review on the subject). This behavior
in the lightcurve was interpreted as evidence for the formation of dust in the nova ejecta (Geisel, Kleinmann, \& Low
1970) and first modelled by Clayton \& Hoyle (1976) and Clayton \& Wickramasinghe (1976). The dust formation
interpretation was challenged by Bode
\& Evans (1981), who suggested that the rise in the IR lightcurve is the result of the delayed reradiation of the
UV-optical light by pre-existing circumstellar dust (an infrared echo). However, the echo model fails to
explain various aspects of the nova development, and the dust formation scenario is now the most commonly
accepted explanation for the IR development in novae (Gehrz 1988). 

Novae are not likely to be important sources of interstellar dust. Novae are about 4000 times more frequent than Type Ia SNe
(van den Bergh \& Tamman 1991). With a local SN Ia rate of 3.7 10$^{-3}$ pc$^{-2}$ Gyr$^{-1}$ (see Table 2), the typical
nova rate is  $\sim$ 15 pc$^{-2}$ Gyr$^{-1}$. Most novae produce carbon dust. Dust masses in the novae ejecta range from 
$\sim 2\ 10^{-8}$ to $\sim 4\ 10^{-7}$ \msun (Gehrz 1988). Adopting an average value of about $\sim 2\ 10^{-7}$, we get that
the dust production rate of novae is $\sim 3\ 10^{-6}$ \msun pc$^{-2}$ Gyr$^{-1}$. Silicate production has been observed
only in one nova, which developed an optically thin dust shell. The silicate production yield of novae is therefore smaller
than that of carbon. In spite of their negligible contribution to the total dust production rate, novae condensates may be
important sources of some isotopic anomalies in meteorites.

\subsubsection{Following Explosive Conditions: Type II Supernovae}

Supernovae have been suggested as important sources of interstellar dust long before the
existence of any supporting evidence (Cernuschi, Marsicano, \& Codina 1967; Hoyle \& Wickramashinghe 1970).
In a then controversial paper, Clayton (1975) suggested supernovae condensed dust as a
source of isotopic anomalies in $^{129}$I, $^{26}$Al, and $^{244}$Pu isotopes found in meteorites.
This view has since then gained wide acceptance, and SN-condensed dust is now accepted
as the origin of a wide range of r- and p-process anomalies in meteorites (see review by Zinner 1996).

Various observational
tests for the formation of dust in SNe were suggested. They fall basically into three
categories: (1) searches for the development of an infrared excess in the SN light curve, 
similarly to that observed
in novae (Aiello, Bonetti, \& Mencaraglia 1974; Falk, Lattimer, \& Margolis 1977); (2) searches for evidence of
extinction of background stars or nebular emission in the SN ejecta (Trimble (1977); and (3) searches for the IR
emission from dust in the ejecta of young unmixed SNRs (Dwek \& Werner 1981, Dwek \& Arendt 1992).

SN 1979c and SN 1980k were the first SNe that were observed at near-IR wavelength and
developed an IR excess about 7 - 9 months after their explosion (Dwek et al.
1983; Dwek 1983). The IR excess from SN 1979c could only be explained
as an IR echo, since the IR luminosity was in excess of the extrapolated bolometric luminosity 
at the time the excess appeared (Dwek 1983), leaving only SN1980k as a possible site of dust formation.
Attributed to an echo, the IR excess from both SNe was used to infer the mass loss rates for the progenitor stars
(Dwek 1983), and found to be consistent with those required to explain the radio light curve from these SNe. 
Consistent with an echo interpretation, the IR observations of SN1980k provided therefore at most ambiguous evidence for
the formation of dust in SN ejecta.

The explosion of SN 1987A provided the first direct and unambiguous evidence for the formation
of dust in SN ejecta, as well as a new set observational criteria for the establishment of
this fact.
The first hint that dust was about to form in SN 1987A came from the detection of CO in its ejecta
(Danziger et al 1991). The formation of CO molecules has in many cases been a precursor to the imminent
formation of dust in novae (Gehrz et al. 1980). In addition, observers found also evidence for the presence of SiO
molecules in the  ejecta of SN 1987A (Roche, Aitken, \& Smith 1991).  

The formation of dust in SN 1987A occured around day 530 after the explosion, and manifested itself in many different,
and sometimes complimentary, ways: (1) in the evolution of a near- and mid-IR excess in the SN lightcurve that occurred
concurrently with a drop in the UV-visual output from the SN (e.g. Bautista et al. 1995, and references therein). The
total energy balance from the SN strongly favored a dust-formation scenario instead of an IR echo (Moseley et al. 1989;
Dwek et al. 1992; Whitelock et al. 1989; Wooden et al. 1993); (2) in the evolution of the emission line profiles of various
elements in the SN ejecta.  Prior to the dust formation epoch these lines exhibited blue- and red-shifted profiles,
characteristic of emission from a spherically symmetric expanding shell. Concurrent with the drop in the UV-visual
output of the SN, line profiles developed an asymmetry, characterized by the diminution of the emission from the
redshifted wing of the profile (Danziger et al. 1991). This line asymmetry persisted even on days 1862 and 2210 after
the explosion (Wang et al. 1996). Absorption by intervening dust within the ejecta is the most logical explanation
for this effect; (3) in the temporal decrease in the emission from various optical lines, such as Mg I]
$\lambda4571$ \AA\ and [O I] $\lambda$ 6300,63 \AA. This drop was attributed to extinction by the newly-formed dust
in the ejecta (Lucy  et al. 1991); (4) in the depletion of various refractory elements from the gas phase. The drop in
the IR line emission from [Si I] 1.65 $\mu$m (Lucy et al. 1991), or [Fe II] 26 $\mu$m (Dwek et al. 1992; Colgan et al.
1994) cannot be attributed to dust extinction, but rather to the possible precipitation of these elements from the
ejecta.

SN condensates were detected in the ejecta of the Crab nebula. Optical continuum images of the Crab (Fesen \&
Blair 1990) revealed numerous dark spots across the synchrotron nebula. The spectral properties and positions of these
spots suggest that they represent dust extinction features instead of synchrotron emission holes in the filaments.
A similar conclusion was reached by Hester et al. (1990) who attributed small-scale "red" features in the IR-to
optical continuum ratio map to extinction by dust.

SN-condensed dust can also be detected from its IR emission. Specifically, dust present in
the evaporative or turbulently-mixed interfaces of clumpy ejecta in a SNR cavity will be collisionally heated by the
X-ray emitting gas, and can give rise to detectable IR emission (Dwek \& Werner 1981). Infrared observations of Cas A
with the {\it Infrared Space Observatory} ({\it ISO}; Lagage et al. 1996) showed indeed clumpy IR emission from the
remnant that is spatially correlated with the fast moving knots previously detected in optical emission lines
(Fesen, Becker, \& Goodrich 1988). IR 20 $-$ 50 $\mu$m spectra taken from various position in Cas A, confirmed a dust origin
for the emission (Moseley et al. 1997).  

A difficult problem is the
determination of the amount and composition of the dust that formed in the ejecta. Observationally, the 15 $-$ 30 $\mu$m
spectrum of SN 1987A was found to be featureless, and consistent with emission from
either carbon or silicate type dust (Wooden et al. 1993; Moseley et al. 1989; Dwek et al. 1992). In principle, the dust
composition can also be inferred from the selective extinction of the various emission lines in the SN ejecta.
Furthermore, the evolution of the extinction with time can be used to determine changes in the dust composition and
size distribution during the condensation period. Unfortunately, the quality of the data does not allow for such
detailed analysis of the observations (Lucy et al. 1991).

Theoretically, the calculated composition of the dust depends crucially on the assumptions made regarding the degree
and type of mixing in the SN ejecta.  Lattimer, Schramm,
\& Grossman (1978) provided the first estimates of SN dust compositions by applying chemical equilibrium
condensation calculations to various zones of an "onionskin" presupernova model. They did not consider the effect of
mixing between the various zones of the ejecta. So in their calculations, zones were characterized by their C/O ratio,
and depending on its value, either Ca- Ti- and Al-rich grains, or C-rich grains such as TiC, SiC, Fe$_3$C or CaS would
form in the ejecta. Clayton \& Ramadurai (1977) suggested that SN ejecta may contain
various sulfide particles, such as TiS, MgS, or FeS, a prediction that may have been confirmed by the decrease
in the IR line emission from Fe and S in SN 1987A after the dust-formation epoch (Dwek et al. 1992, Colgan et al. 1994).
More speculatively, Nuth \& Allen (1992) and Clayton et al. (1995) suggested SN produced diamonds as carriers of nobel
gas anomalies in meteorites.

SN 1987A provided theorist with a wealth of new information to predict the emerging dust
composition. In particular, various lines of evidence suggested that the ejecta of SN 1987A was at least partially
mixed. Consequently, Kozasa, Hasegawa,
\& Nomoto (1989) calculated the emerging grain composition, mass, and size distribution fore a variety of mixing
conditions. In a subsequent paper Kozasa, Hasegawa, \& Nomoto (1991) repeated their calculations under the
assumption that the ejecta was completely mixed. Since the global C/O ratio in the ejecta is
$<$ 1, all the carbon in this model is locked up in CO. Consequently, their results predicted the formation of grains
with composition typical of O-rich environments, such as Al$_2$O$_3$, Fe$_3$O$_4$, and silicate grains. 

All these calculations illustrate the sensitivity of the emerging dust composition on model assumptions, and the
difficulty of predicting the dust composition even in the best studied supernova. It is clear however, that even if
mixing did occur in SN 1987A, that it was macroscopic, rather than microscopic in nature (see \S 3.6.5 below).
As a result, some regions in the ejecta are expected to remain C-rich and form carbon-rich grains. Furthermore, even if
in some regions the C and O are mixed on a microscopic level, not all the carbon is expected to be locked up in CO. This
results from the fact that the same mixing process responsible for the formation of CO can admix He$^+$ into the mixed
layer, destroying the CO by the charge exchange reaction: He$^+$ + CO
$\rightarrow$ He + C$^+$ + O (Liu, Dalgarno, \& Lepp 1992).

The mass of the dust expected to form in SNe is equally uncertain. The IR emission from SN 1987A suggest a dust mass of
$\sim (1-10)\ 10^{-4}$ \msun (Dwek et al. 1992; Wooden et al. 1993). However, the radiating dust may comprise only a
small fraction of the total amount of dust formed in the ejecta. Most of the dust is probably cold and concentrated in
optically thick clumps (Lucy et al. 1991). A more reliable estimate of dust mass can be obtained from the amount of
extinction, or the drop in the line emission of various refractory elements in the SN ejecta. From these observations we
can infer that most of the condensible elements precipitated efficiently from the expanding SN ejecta. If so,
the expected dust mass can range from 0.1 to 1 \msun. 

\subsubsection{Following Explosive Conditions: Type Ia Supernovae}
To date there is no direct evidence for the formation of dust in Type Ia supernovae. There are some differences
between Type Ia and Type II events which may inhibit the formation of dust in the former. Ejecta velocities, and
the abundance of radioactive material are higher in Type Ia compared to Type II events. These may prevent
the formation the formation of dust for reasons similar to those that inhibit the formation of dust in fast
novae (Gehrz 1988). However, Clayton, Arnett, Kane, \& Meyer (1997) presented
convincing arguments in favor of dust production in Type Ia events. 

One group of the interstellar dust particles isolated and studied in meteorites consists of type-X SiC
particles, with isotopic composition characterized by large excesses of $^{44}$Ca (the radioactive decay product
of $^{44}$Ti) , $^{28}$Si, and $^{15}$N. Clayton et al. (1997) argue that the assembly of dust with such
composition in Type II events will require microscopic mixing of the ejecta, transporting various
elements from different zones in the ejecta to the SiC-X dust condensation site. 2-D numerical and laboratory
laser simulations of hydrodynamical explosion show evidence for macroscopic mixing in the ejecta, but not the
kind of microscopic mixing required to produce SiC-X grains.

In light of these difficulties, Clayton et al. (1997) suggested that Type-X SiC grains condense out of SN Ia events
that explode with a cap of He atop their CO structure. The SiC
particles condense out of ejecta that underwent explosive He burning in $^{14}$N-rich matter, and the
relative abundances of the C, Si, and trace N, Mg, and Ca match those of the SiC-X particles without the need
to resort to ad hoc mixing scenarios.

The depletion of Fe in the general ISM is particularly sensitive to the ability for dust to form in SN Ia events. Iron is
observed to be depleted with $\Delta$(Fe) $\gtrsim$ 0.70 in refractory grain cores (Sofia, Cardelli, \& Savage 1994). About one
third of the interstellar iron is produced in SN Ia events. So if Type Ia events do not produce dust, the depletion of Fe
(neglecting grain destruction by SNRs or Fe accretion in clouds) is expected to be $\sim$ 60\% at most. If Type Ia SN {\it do}
produce dust, it will have important consequences for the composition of iron dust in the ISM. Since Fe is produced in large
excess over any other refractory element including oxygen (Thielemann et al. 1993), it will likely condense out in
pure metallic form (see \$ 3.8 below).

\subsection{Dust Production Yields}

We distinguish between stars
with masses below M$_w$ = 8 M$_{\odot}$, and those with masses above this value. For stars with M $\leq$ M$_w$, the
yield of the dust for each mass is simply determined by the C/O ratio in the ejecta. When the C/O ratio larger
than unity, we assume that all the {\it excess} C is locked up in {\it carbon} dust. Its exact composition, whether
graphite or amorphous carbon, is not important for the purpose of this paper.  When the C/O ratio is less than unity,
all the refractory elements: Mg, Si, Ca, Ti, and Fe, and about an equal amount of O {\it by number} are assumed to
condense out of the gas. We refer to this dust as {\it silicate} dust. We will ignore the cases in which C/O
$\approx$ 1, since they are rare, and will not contribute substantially to the abundance of dust in the ISM.

To be more quantitative, we denote by M$_{ej}(A$,M) be the mass of an element $A$ in the stellar ejecta, and by
M$_{dust}(A$,M) the mass of
$A$ that is locked up in dust. Considering the refractory elements: C, O, Mg, Si, S, Ca, Ti, and Fe, we have adopted
the following prescription to calculate the dust composition and production efficiency in the various stellar sources:

\noindent
1) For stars with M $\leq$ M$_w$ = 8 \msun:

\noindent
$\ \ \ \ \ \ $ a) for stellar masses $M$ in which C/O $>$ 1 in the ejecta
\begin{eqnarray}
	M_{dust}(C,M) &  = & \delta_{cond}^w(C)\ \left[M_{ej}(C,M)-{3\over 4}\ M_{ej}(O,M)\right] \nonumber \\
 M_{dust}(A,M) & = & 0 \ \ \ \ \ for\ \  A = \{\ O,\ Mg,\ Si,\ S,\ Ca,\ Ti,\ and\ Fe\} 
\end{eqnarray}
$\ \ \ \ \ \ $ b) for stellar masses $M$ in which C/O $<$ 1 in the ejecta
\begin{eqnarray}
	M_{dust}(C,M) &  = & 0 \nonumber \\
 M_{dust}(A,M) & = & \delta_{cond}^w(A)\ M_{ej}(A,M)\ \ \ \ \ \ for\ \  A = \{\ Mg,\ Si,\ S,\ Ca,\ Ti,\ and\ Fe\} \\
 M_{dust}(O,M) & = & 16 \sum_{A=\{\ Mg,\ Si,\ S,\ Ca,\ Ti,\ Fe\}}\delta_{cond}^w(A)\ M_{ej}(A,M)/\mu(A) \nonumber  
\end{eqnarray}

\noindent
where $\mu(A)$ is the mass of the condensing specie in amu.

\noindent
2) For stars with M $>$ M$_w$ = 8 \msun:

\noindent
\begin{eqnarray}
	M_{dust}(C,M) &  = &  \delta_{cond}^{II}(C)\ M_{ej}(C,M)  \nonumber \\
 M_{dust}(A,M) & = & \delta_{cond}^{II}(A)\ M_{ej}(A,M)\ \ \ \ \ \ for\ \  A = \{\ Mg,\ Si,\ S,\ Ca,\ Ti,\ and\ Fe\} \\
 M_{dust}(O,M) & = & 16 \sum_{A=\{\ Mg,\ Si,\ S,\ Ca,\ Ti,\ Fe\}}\delta_{cond}^{II}(A)\ M_{ej}(A,M)/\mu(A) \nonumber
\end{eqnarray}

\noindent
An equation identical to equation (25) was
used for Type Ia SN, but with the $\delta_{cond}^{II}(A)$'s replaced with $\delta_{cond}^{I}(A)$'s.

The above equations assume that in stars with masses $< M_w$, the C and O in the ejecta are microscopically mixed so
that the maximum possible amount of CO is formed. For higher mass stars and Type Ia SNe, the ejecta was assumed to
be only macroscopically mixed, allowing for the formation of both, silicate and carbon type grains. In principle,
we could have followed the detailed composition of the dust returned by each stellar mass into the ISM by choosing
the right combination of $\delta_{cond}$ for each element. In practice, we chose $\delta_{cond}^w = 1$, and
$\delta_{cond}(A)^{II} = \delta_{cond}(A)^I$ = 0.8 for \{A\} = \{ Mg, Si, S, Ca, Ti, Fe \}, and equal to 0.5 for carbon.
 The $\delta_{cond} <$ 1
 values for Type II and Ia SNe presumably take into account the possible incomplete condensation of the elements in these
objects, and the possible destruction of SN condensates during their injection into the ISM (see \S 6.1 below). The
choice of the $\delta_{cond}$'s for these objects is quite arbitrary.

All elements other than carbon are considered to be "silicate" grains, that is
M$_{silicates}=\sum_{\{  A\
\neq\ C\}}M_{dust}(A)$, although it is clear that in some cases sulfides or other grain composition may emerge from the
ejecta. A particular case in point is the composition expected from Type Ia SNe. The models of Nomoto and coworkers
give the following masses (in
\msun) for the nucleosynthetic products from these objects: \{C, O, Mg, Si, S, Ca, Fe\} = \{0.020, 0.130, 0.016, 0.150,
0.083, 0.015, 0.742\}, where we omitted the noble gases. Clearly, Type Ia SNe produce more iron than can be bound with
O or S to form various oxides or sulfides. Much of the iron from Type Ia SNe may therefore condense as pure iron.
Nitrogen is neither expected to condense out in stellar or SN ejecta nor is it expected to accrete onto grains in the ISM
(Barlow 1978c). We therefore ignored the evolution of N in the dust phase of the ISM.

\subsection{Relative Production Yields by the Various Sources}
Taking the C/O ratio as a
criterion for the dust composition, the range of stars that form carbon dust depends on the initial stellar
metallicity, as well as details of the dredgeÐup process (Renzini \& Voli 1981; Frost \& Lattanzio 1996 for a review).
This is illustrated in Figure 6, which depicts the C and O production yield of stars in the 1 $-$ 40 M$_{\odot}$ mass
range, as well as the C/O ratio in the ejecta for stars with an initial metallicities of Z = 0.0040 (Fig 6a) and Z =
Z$_{\odot}$ = 0.020 (Figure 6b). When the initial stellar metallicity is low, the amount of carbon needed to be dredged up
to the surface in order to increase the surface C/O ratio to a value larger than unity is low. Consequently, the lower mass
cutoff of stars that become carbon stars is 1.7 M$_{\odot}$ at $Z$ = 0.0040, and somewhat higher, 2.3 M$_{\odot}$, at
$Z$ = 0.020. The upper mass cutoff on stars that become carbon stars is determined by the operation of the hot bottom
burning process (HBB). In this process, the convective carbon rich envelope reaches down into the H-burning shell, and it
bottom becomes hot enough for the CN cycle to process the carbon that was previously convected to the surface, into
nitrogen. Stellar yields with 
$\alpha=3/2$ in the RV81 tables include the effect of the HBB process, and the high mass cutoff for stars
that become C-stars is 6.5 M$_{\odot}$ at $Z$ = 0.0040, and 5.3 M$_{\odot}$ at $Z$ = 0.020. Observationally, the
initial mass range of main sequence stars that will become C-stars depends on the sample selection
criteria. Jura (1991) finds a progenitor mass range between 1 and 5
\msun from observations of high-luminosity C-stars. Barnbaum et al. (1991) finds initial mass ranges between 2.5 and 4
\msun from an analysis C-stars with expansion velocities above $\sim$ 17.5 km s$^{-1}$. Kastner et al. (1993) find
that progenitor masses are at least 2 \msun from a sample of distant low Galactic latitude AGB stars, and Groenewegen,
van den Hoek, \& de Jong (1995) derive a mass range of $\sim$ 1.5 to 4 \msun from a comparison of observations with
synthetic AGB evolution calculations. These values are similar to the theoretically derived values of RV81, and the
differences should be a measure of the uncertainties in the calculated carbon dust abundance when compared with
observations. For stars with masses above 8 M$_{\odot}$ we allow, for reasons previously discussed, the formation of both,
silicate and carbon dust.

Figure 7 depicts the SMS$-$weighted carbon and silicate dust yield of the stars in the 1 $-$ 40 \msun mass range with
initial metallicities of Z = 0.0040 (Fig 7a) and Z = 0.020 (Fig 7b), as a function of the initial stellar mass.  The
figure shows that carbon grains are predominantly produced in stars with masses in the 2 $-$ 5 \msun range, whereas
silicates are predominantly produced in higher mass stars. This fact has important observational consequences,
since low mass stars return their ejecta to the ISM a considerable time after their formation (see \S8.5).

%******************************
\section{GRAIN DESTRUCTION}
%******************************

\subsection{Destruction During the Injection into the ISM}
Circumstellar dust grains formed by quiescent mass loss may be altered during their injection phase into the
ISM. They may be shattered, coagulate, or grow by accretion in the stellar outflow, but are mostly expected
to survive the injection phase. This is not the case with SN condensates, which are injected at velocities
exceeding 1,000 km s$^{-1}$ into a a supernova cavity, containing shock-heated circumstellar/interstellar gas.
{\it ISO} infrared images of the knots in Cas A vividly illustrate the perils supernova condensates can be
subjected to after their formation. The clumpy ejecta containing the SN-condensed dust can evaporate or be
turbulently mixed into the hot SN cavity gas where they will be subjected to destruction by thermal sputtering.
Correlated SN may minimize the destruction of {\it interstellar} dust (McKee 1989), but may still be efficient in
destroying newly-formed SN condensates. This aspect of grain destruction was previously overlooked, and can be
incorporated into evolutionary models either by instantaneously updating the local metallicity of the ISM
with the explosion of each SN, and calculating the effect of subsequent SNe on this enriched parcel of gas; or by
reducing the condensation efficiency in the SN ejecta to take the destruction probability into account.
A fraction of the SN condensates can escape destruction if the SN ejecta is mixed with the cold ISM before it evaporates
in the SN cavity, or before subsequent SNe explode in its vicinity. The kinetic energy of the clump is then transferred
via a cascade of turbulences to the ambient medium. The SN condensates may then just be subjected to shattering and
coagulation processes,which will preserve the mass of refractory elements in the dust, instead of the thermal
sputtering process in the SN cavity.

\subsection{Destruction by Expanding Supernova Blast Waves}
In the ISM, grains can be destroyed by thermal sputtering, evaporation in grain-grain collisions, thermal
sublimation, and chemical sputtering. Of these processes, grain destruction by SNRs is the most important
mechanism for cycling the dust back to the gas phase. Let m$_{dest}$(A,r,t) be the total mass of element
$A$ that, initially locked up in dust, is returned to the gas  by a single SNR expanding at location $r$ and time
$t$ during its evolutionary lifetime. The rate at which $A$ is returned to the gas is then given by:

\begin{equation}
\left[{d\sigma_{dust}(A,r,t)\over dt}\right]_{SNR} = m_{dest}(A,r,t)\ {\cal R}_{SN}(r,t)
\end{equation}

\noindent
where ${\cal R}_{SN} = {\cal R}_{SNIa} + {\cal R}_{SNII}$, is the combined rate of Type Ia and Type II events in the
Galaxy in units of pc$^{-2}$ Gyr$^{-1}$.
Equation (26) can also be written in the form:

\begin{equation}
\left[ {d\sigma_{dust}(A,r,t)\over dt} \right] _{SNR} \ =\  \sigma_{dust}(A,r,t) \left[{m_{dest}(A,r,t)\over
\sigma_{dust}(A,r,t)}\right] 
                                                   {\cal R}_{SNR}(A,r,t)
   \ \equiv \ {\sigma_{dust}(A,r,t)\over \tau_{SNR}(r,t)}
\end{equation}

\noindent
which defines the lifetime of element $A$ against destruction by SNRs (Dwek \& Scalo 1980; McKee 1989). 

The value of $\tau_{SNR}$ plays a key role in calculating the interstellar depletion of element $A$. 
Calculations of $\tau_{SNR}$ involve knowledge of the physics of grain destruction by thermal sputtering and
grain-grain collisions (Barlow 1978a,b; Draine \& Salpeter 1979a;
Jones et al. 1994; Tielens et al. 1994; and Borkowski \& Dwek 1995), as well as the application of these grain
destruction rates to interstellar shocks and SNRs (Shull 1977; Draine \& Salpeter 1979b; Dwek \& Scalo 1980; 
Jones et al 1996; McKee et al 1987; Seab 1987). In particlular, McKee (1989), stressed the importance of the
effect of correlated SNRs on calculating the grain lifetime. Dust evaporation by grain-grain collisions is the
most difficult to model, yet becomes the dominant grain destruction mechanism during the radiative phase of
remnant evolution, when most of the ISM is swept up. Grain destruction efficiencies by radiative shocks have been
recently modeled by Jones et al. (1994), and global refractory grain lifetimes of $\sim$ 0.4 and 0.22 Gyr have been estimated,
respectively, for carbonaceous and silicate grains (Jones et al. 1996).

Grain destruction lifetimes are also expected to vary over the lifetime of the Galaxy. To explore this time dependency, we
write m$_{dest}$, the mass of dust destroyed by a single SNR as:

\begin{equation}
m_{dest}(A,r,t) = \left({\sigma_{dust}(A,r,t)\over \sigma_{ISM}(A,r,t)}\right) \epsilon M_{SNR}
\end{equation}

\noindent
where M$_{SNR}$ is the total mass of ISM gas swept up by the remnant during its lifetime, and $\epsilon$ is an average grain
destruction efficiency. 
The grain destruction lifetime is then:

\begin{equation}
\tau_{SNR}(A,r,t) =  \left(\epsilon\ M_{SNR}\right)^{-1}\ \left[{\sigma_{ISM}(r,t)\over {\cal R}_{SN}(r,t)}\right]
\end{equation}

\noindent
The amount of ISM mass swept up at a given velocity by an isolated SNR, does not depend on the ambient density, a simple
consequence of momentum conservation during the late stages of its evolution. If we
neglect the weak dependency of $\epsilon$ on the ISM density (Jones et al. 1996), then the quantity
$\epsilon\ M_{SNR}$ can be assumed to be constant with
time. With this approximation, the lifetime of element $A$ against destruction by SNRs can be written in terms of 
$\tau_{SNR}(A,(R_{\odot}, t_G)$, the current lifetime in the solar circle, as:
 
\begin{equation}
\tau_{SNR}(A, r, t) = \left[{\sigma_{ISM}(r,t)\over \sigma_{ISM}(R_{\odot}, t_G)}\right]
                      \left[ { {\cal B}(r,t)\over {\cal B}(R_{\odot}, t_G)}\right]^{-1}\ \tau_{SNR}(A,(R_{\odot}, t_G)
\end{equation}

\noindent
where we assumed that the SMS is spatially and temporally constant so that ${\cal R}_{SN} \propto {\cal B}$.
Figure 8 depicts the the value of $\tau_{SNR}$ at the Galactic center and in the solar circle as a
function of time.   

\subsection{Grain Destruction in the Starburst Galaxy M82}
It is interesting to compare the theoretically derived grain destruction rates and timescales in our Galaxy (Jones et al.
1996) with those "observed" in external galaxies. An empirical grain destruction rate can be derived from [Fe II] 1.644\mic
$^4$D$\rightarrow  ^4$F line transition in external galaxies. Fe has a first ionization
potential of 7.87 eV, and is therefore easily ionized by the interstellar radiation field, provided it is not
locked up in dust. Since Fe is expected to be depleted into dust, Greenhouse et al.
(1991) suggested that the [Fe II] emission can be used as a tracer
for supernova remnants, since they are the primary agents for grain destruction in the ISM.
Fabry-Perot [Fe II] 1.644\mic line images of M82 confirmed this idea (Greenhouse et al 1997). The
observations showed that the  [Fe II] emission is spatially correlated with known
radio SNR in that galaxy (and not with H II regions). Their observations can be used to estimate the destruction
lifetime of Fe in that galaxy. 

\noindent
To do so, we rewrite equation (26) for a galaxy as a whole in the form:
\begin{equation}
\tau_{SNR}(A) = \left({Z_{dust}(A)\over \dot {\cal N}_{SN}}\right)\ \left({m_{dest}(A)\over
M_{ISM}}\right)^{-1}
\end{equation}

\noindent
where $\dot {\cal N}_{SN}$ is the SN rate in yr$^{-1}$, and M$_{ISM}$ is the total ISM mass in the galaxy.
In using equation (30), we will assume that the Fe$^+$ observed in the 1.644 \mic emission line represents the mass of iron
that is returned by SNR to the gas phase, and that all the remaining iron is locked up in dust. A lower limit to the
mass of Fe$^+$ can be derived by assuming that after the emission of the 1.64 $\mu$m photon all the Fe$^+$ is
instantaneously recycled back to the excited level. The observations of Greenhouse et al. give an
extinction corrected [Fe II] 1.644
\mic luminosity of
$1.1\ 10^{40}\ erg\ s^{-1}$ for an adopted distance of 3.2 Mpc. The $A$ coefficient for the transition is
$4.65\ 10^{-3}\ s^{-1}$ (Nussbaumer \ Storey 1988), giving a total mass of $\sim$ 0.083 \msun of Fe$^+$ in
the galaxy. A more realistic assumption is that the Fe$^+$ is in LTE, in which case only $\sim$ 10\% of the ions will
be in the excited level, giving an Fe$^+$ mass of $\sim$ 0.83 \msun. However, LTE requires electron densities
$\gtrsim 10^5$ cm$^{-3}$, which are not likely to be met in the general ISM. For more realistic densities of $\sim \ 
10^2$ cm$^{-3}$, the relative population in the excited level is only $\sim$ 10$^{-4}$, giving an  Fe$^+$ mass of $\sim$
800 \msun. The ISM mass in the inner 6 kpc of M82 is estimated to be
$\sim 2\ 10^8$ \msun in H I (Yun, Ho,
\& Lo 1993), and 
$\sim 2\ 10^9$ \msun in H$_2$ (Young \& Scoville 1984). In deriving the H$_2$ mass Young \& Scoville (1984) adopted
a Galactic N(H$_2$)/N(CO) conversion factor, which may an overestimate for M82 (Wild et al. 1992; Lord et al. 1996).
Assuming a solar Fe abundance of
$1.28\ 10^{-3}$, a somewhat lower ISM mass of 10$^9$ \msun, and a SN rate of
${\cal N}_{SN} \sim\ 0.10\ yr^{-1}$ (Kronberg, Biermann, \& Schwab 1981), we get that

\begin{equation}
\tau_{SNR}(Fe) \approx  \left( {1.28\ 10^{-3}\over 0.10}\right)\ \left({800\over 10^9}\right)^{-1} \approx  
    2\ 10^4\ yr  
\end{equation}

\noindent
This value assumes that all the iron was initially depleted onto dust grains. However, even if only 10\% of the iron was
initially in the dust, so that, $\tau_{SNR}(Fe) \approx 2\ 10^5$ yr, this value would still be significantly smaller than all
current theoretical estimates, and significantly smaller that typical lifetimes of starbursts ($\sim 100\ 10^6$ yr). We
therefore must conclude, that the starburst region itself will end up mostly dust$-$free, and that a significant fraction
 of the infrared emission observed from starbursts is stellar radiation that is reprocessed by a foreground screen of dust.  

%***************************************************
\section{GRAIN GROWTH IN THE INTERSTELLAR MEDIUM}
%***************************************************
In dense interstellar clouds grains can grow by the accretion of condensible elements onto {\it preexisting}
refractory cores. For simplicity we will assume that the dust particles consist of one type element $\bar A$,
with a mass of $\mu_A$. Let n$_g(A)$ be the number density of $A$ in the {\it gas} phase of the
ISM. The rate per unit volume at which the number of atoms, N$_A$, in the dust grows by accretion is given
by:

\begin{equation}
{dN_A\over dt} = \alpha\ \pi a^2\ n_g(A)\ n_{gr}\ \bar v
\end{equation}

\noindent
where $\alpha$ is the sticking coefficient of element $A$ to the grain, $a$ is the grain radius, n$_{gr}$ is
the number number density of dust grains, and
$\bar v = (8kT/\pi m_A)^{1/2}$ is the mean thermal speed of $A$ in the gas. The rate at which the mass
of the ISM dust phase grows by accretion can be written in terms of global Galactic quantities as:

\begin{equation}
{d\sigma_{dust}(A)\over dt} = \alpha\ \left({\pi a^2\over m_{gr}}\right) \mu_A \ n_g(A)\ \bar v\ 
                             \sigma_{dust}
\end{equation}

\noindent
where m$_{gr}$ is the mass of an individual dust grain. Elemental accretion takes place in high density
molecular clouds (n$_H \gtrsim 10^3\ cm^{-3}$). The mass density of the element
$A$ can then be written in terms of the cloud density, $n_c$, as: $\mu_A \ n_g(A) = 2m_H\
n_c\ (1-\sigma_{dust}(A)/\sigma_{ISM})$. Equation (21) can be rewritten in the form:

\begin{eqnarray}
{d\sigma_{dust}(A)\over dt} & = & {\sigma_{dust}(A)\over \tau_{accr}(A)} \nonumber \\
                         & = & {[1 - \Delta(A)]\over \tau_0(A)}\ \sigma_{dust}
\end{eqnarray}
where the depletion factor $\Delta(A) = \sigma_{dust}(A)/\sigma_{ISM}(A)$ is fraction of the element $A$ that is locked
up in the dust. The accretion timescale of element $A$ onto dust grains is then

\begin{equation}
\tau_{accr(A)} \equiv \tau_0/\left[1-\Delta(A)\right] 
\end{equation}

\noindent
where $\tau_0(A)$ given by:

\begin{equation}
\tau_0(A)^{-1} \equiv  \alpha\ \left({\pi a^2\over m_{gr}}\right) \mu n_c\ \bar v\
\end{equation}

\noindent
where $\mu$ is the mean molecular weight of the gas.
The timescale $\tau_0$ can be estimated by taking typical numbers for the various dust and cloud properties that
determine the accretion timescale. For 0.1 $\mu$m radius grains with mass densities of $\sim$ 3 gr cm$^{-3}$,
 an accreting refractory of mass  $\mu_A$ = 20 m$_H$, molecular cloud densities of $\sim 10^3$ cm$^{-3}$,and
temperatures of $\sim $ 20 K, we get a value of
$\tau_0 \approx 3\ 10^4$ yr. For $\Delta \approx$ 0.7, that is, 70\% of the element $A$ is in the dust phase, we
get $\tau_{accr} \approx 10^{5}$ yr for the accretion time. This timescale is short compared to the typical
lifetime of a molecular cloud. From an extensive CO survey of regions around open star clusters, Leisawitz, Bash,
\& Thaddeus (1989) show that clusters with ages above $\sim 10^7$ yr retain only low mass molecular clouds in
their vicinity. The smaller remaining pieces of the cloud complexes may have a longer lifetime, about
(1-2) 10$^7$ yr (Larson 1987). Since the abundance of molecules is not zero in molecular
clouds, this implies the existence of various mechanisms that eject newly accreted material back to the gas phase.
Such mechanisms include normal evaporation following some cosmic-ray or UV induced stochastic heating event,
explosive desorption, or grain-grain collisions (see Jenniskens et al. 1993 and references therein). The effects
of these mechanisms were simulated in the laboratory by Jenniskens et al. (1993). In their experiments they
exposed grain material to photons and ions in order to simulate the formation and evolution of grain mantles on
interstellar grains. They estimate that interstellar dust particles accrete a layer of 200 \AA\ during their
lifetime in a molecular cloud. For initial grain radii of
$\sim$ 0.05 $\mu$m, these results imply that the dust mass increased by a factor of $\sim$ 2.7, an e-folding time.
The effective accretion timescale is therefore equal to $\sim 6\ 10^7$ yr, the lifetime they adopted for a typical
molecular cloud.

However, at any given time, only a fraction of the total interstellar dust population resides in dense molecular
clouds and is growing by accretion. To be applicable to equation (6) which describes the dust evolution averaged over the
ISM phases, the effective accretion timescale derived
above must be divided by the fraction of dust particles in molecular clouds. If we assume that this fraction is
about 1/2, then the ISM-phase averaged accretion timescale is $\sim (1-2)\ 10^8$ yr, or about one-half the grain destruction
time by SNRs.

The accretion timescale is not expected to be constant over the lifetime of the Galaxy. In our global approach,
which consists of a one$-$phase ISM, the time dependence of $\tau_{accr}$ originates primarily from the
changing mass fraction of molecular clouds in the Galaxy. The value of $\tau_{accr}$ derived for molecular clouds must
therefore be divided by the fraction of the ISM mass that is in molecular clouds, that is, the ratio
$\sigma_{MC}$/$\sigma_{ISM}$. We adopted a complicated prescription for the dependency of the stellar birthrate on the ISM
and stellar mass densities (see eq. 15). However, if we assume that it is simply a linear function of the mass density of
{\it molecular} material, then ${\cal B} \propto \sigma_{MC}$, giving
$\tau_{accr} \propto \sigma_{ISM}$/${\cal B}$. The attractive feature of this simplifying assumption is that the 
grain accretion and destruction timescales evolve identically in time.
 
Figure 8 depicts the accretion timescale as a function of time. Since $\tau_{SNR}$
and $\tau_{accr}$ both scale with time as $\sigma_{ISM}(r,t)$/${\cal B}(r,t)$, the current adopted $\tau_{accr}$/$\tau_{SNR}$
ratio of ${1 \over 2}$ is constant with time. Table 1 lists the current grain destruction and accretion timescales used in this
paper.

%****************************
\section{MODEL RESULTS}
%****************************

We calculated the evolution of the disk, the ISM, the elemental and dust abundances by using a simple fourth$-$order
Runge$-$Kutta integration procedure. Nucleosynthesis yields depend on the initial stellar metallicities, Z$_*$, 
and are therefore time dependent. 
To facilitate the numerical computations, we first calculated the evolution of the total
metallicity, $Z_{ISM}$, using constant nucleosynthesis yields, obtained by averaging the yields at
 Z$_*$ = 0.0040 and 0.020. 
This gave
us a first order solution for the evolution of $Z_{ISM}$ with time. Nucleosynthesis and dust production yields, 
given as a function Z$_*$, were then mapped onto a time grid using the calculated $Z_{ISM}\ -\ t$ relation. The 
delayed recycling effect plays an important role in determining the evolution of the dust composition, and to 
ensure a smooth solution,
we interpolated the stellar yields on a fine grid of 70 mass points in the 1 - 40 \msun mass interval. 
The results of our
calculations are presented below.  

\subsection{Dust Sources and Sinks, and the Roles of Supernovae and Supernova Remnants}

Figure 9  compares the dust production rate of various sources: Type II and Type Ia SNe, and low mass stars, to the 
destruction
rate by astration. The various panels depict the silicate and carbon production rates at the Galactic
center and in the solar circle. Values at the current epoch in the  solar circle are listed
in Tables 3 and 4. The
tables shows that supernovae are the most important source of interstellar dust, contrary to the estimate of Gehrz 
(1989). The
table shows considerable agreement between the production rates resulting from our detailed calculations and the 
observational values summarized by Jones \& Tielens (1994). However, we find that novae dust production rates have been 
significantly overestimated in previous work. The tables also illustrates the role of
supernovae and their remnants as, respectively, interstellar dust sources and sinks at the current epoch. Grain
destruction rates are $\sim$ 50 and $\sim$ 400 M$_{\odot}$ pc$^{-2}$ Gyr$^{-1}$ for, respectively, carbon and silicate grains,
whereas the production rate by both, Type Ia and II SN is $\sim$ 1.5 and $\sim$ 10 M$_{\odot}$ pc$^{-2}$ Gyr$^{-1}$ for carbon
and silicate grains, respectively. Supernovae are therefore net destroyers of interstellar dust {\it at the current epoch}.
However, at earlier epochs, when the ISM metallicities are below $Z\ \approx\ Z_{\odot}/40$, they inject more dust into the ISM
than they destroy during the remnant stage of their evolution. The first generations of SNe in the universe play therefore an
important role in the initial dust enrichment of the intercluster medium (Loeb \& Haiman 1997) and of damped Ly$\alpha$
systems (Pei, Fall, \& Bechtold 1991).

\subsection{The Origin of the Elemental Depletion Pattern}
An important test for dust evolution models is their ability to reproduce the elemental depletion
pattern. Figure 10 depicts the fractional abundance of the major refractory elements that is locked up
in dust for three cases. In the first case we assumed that grain destruction
by SNR is {\it exactly} balanced by the rate of grain growth in molecular clouds. This is equivalent to
setting the sixth, seventh and eighth terms in equation (6) to zero. The resulting depletion pattern (open
squares) should then reflect the condensation efficiency in the various sources {\it and} the elemental
dilution factor. This latter effect results from the fact that it is impossible to condense {\it all}
the elements from the gas in {\it all} the sources, even if their condensation efficiency is unity. For
example, even if carbon condenses out with a 100\% efficiency in carbon stars, all other refractory elements, as well as the
carbon locked up in CO molecules, are ejected back into the ISM in gaseous form. On the other hand, O$-$rich stars
eject all the carbon back into the ISM in gaseous form. The absolute level of the depletion also depends on the elemental
condensation efficiencies in Type Ia and Type II supernovae. In our calculations we adopted values less than unity for these
efficiencies. Had we adopted values of
$\delta_I$ or
$\delta_{II}$ = 1, the depletions would be closer to unity. We also did not attempt to introduce
different condensation efficiencies for the different elements, thereby simulating a condensation temperature
dependent nucleation process. 
With so many adjustable model parameters we don't believe that such fine$-$tuning of the model is warranted at
this time.
The depletion trend looks therefore relatively flat for all elements heavier
Mg. 

In the second case we examined the maximum attainable depletions allowing for grain destruction, but
not for grain growth by
 accretion, in the ISM (open diamonds). In the absence of any UV, cosmic ray, or shock processing of the accreted mantle into 
more refractory material, these depletions should reflect the maximum attainable depletions in refractory grain cores,.

In the third case we allowed grains to grow by accretion in molecular clouds (filled circles). The
resulting depletions should be significantly higher than the previous case, since they now reflect core
$+$ mantle depletions. The accretion rates we adopted are somewhat higher that the grain destruction
rates, however, they are limited by the amount of refractory elements available in the gas phase. Accretion rates
 balance the
grain destruction rates by SNRs only when the depletions are about 50\%. Consequently, the resulting depletions are somewhat
lower than those in which all grain processing in the ISM was ignored (open squares). The depletion pattern is relatively flat
above Mg, since we did consider any variations in the accretion timescales for the different elements. We point out 
that accretion caused a significant increase in the depletion of oxygen. In the absence of accretion, the amount of 
oxygen that can be incorporated in the dust is limited by the availability of other metals (Si, Mg, Fe) that it can chemically 
bind to in the gas. This constraint on the condensation efficiency of oxygen in expressed in equations (23) and (24).
These constraints are removed in the accretion process, since oxygen can accrete onto dust as water ice.  

Table 5 compares the calculated core and core+mantle depletions with
observed depletions  summarized by Savage \&
Sembach (1996; their Table 7). A major uncertainty in determining the observed depletions is the set of
reference abundances that should be used for the comparison. This uncertainty is reflected in the range of observed
values quoted in the Table. The table shows that our calculated core + mantle depletions fall somewhat below the
observed values. To get values that are essentially unity, we will have to assume unit condensation efficiency in the
various sources, and that the accretion goes to completion in the molecular clouds. The table also shows that core depletions
are significantly lower than the observed values. Since they are predominantly affected by the grain destruction efficiency
in the ISM, we examined the dependence of these depletions on the grain destruction timescale.

The results of these calculations are shown in Figure 11, where we plot the core depletions of C, Si, and Fe versus the grain
destruction lifetime by SNRs. The grain destruction lifetimes are normalized to the nominal values suggested by Jones et al.
(1996): 0.4, and 0.22 Gyr for carbonaceous and silicate dust, and we adopted a value of 0.3 Gyr for iron dust. Also shown in
the figure are the observed core depletions tabulated by Savage \& Sembach (1996). The two entries given for each element
(there is no data for C), correspond to the different results one obtains by using the sun or B-stars as the reference source
for the undepleted abundances. The figure shows that the nominal grain destruction lifetimes ($\tau_{SNR}$/$\tau_o$ = 1) are
too short to explain the high observed core depletions of Si and Fe. Therefore, {\it a significant fraction of the refractory
grain core must actually be accreted mantle material that has been processed either by UV radiation, cosmic rays, or
interstellar shocks, into more refractory grain material}.

\subsection{Evolution of the Dust Abundance}
The dust abundance and composition plays an important role in determining the Galactic extinction law.
Any evolution in these quantities will therefore result in a time varying Galactic opacity. This effect
will have important consequences for determining the extinction corrections in external galaxies
at various redshifts. Figures 12 and 13 depict the evolution of the ISM (gas + dust) and dust metallicity as a
function of time at the Galactic center and in the solar circle, respectively. Also shown in these
figures are the ratio of the two quantities, and the separate evolution of silicate and carbon grains.

At the Galactic center, the star formation rate peaks at 0.6 Gyr, and slowly declines thereafter. The
metallicity rises accordingly rapidly until t $\approx$ 1 Gyr, after which it rises  more slowly to a value of 
0.03 at the
current epoch. The dust abundance follows that of the metallicity very closely, with Z$_{dust}$/Z$_{ISM}
\approx$ 0.4 throughout most of its evolution. 

At the solar circle, the star formation rate peaks at a later epoch, and the metallicity of the dust
and of the ISM are still rising. As at the Galactic center, the dust abundance follows that of the
metallicity very closely, with Z$_{dust}$/Z$_{ISM} \approx$ 0.4 throughout most of its evolution. This
value is in close agreement with the observational value of
$\approx$ 0.008/0.02 = 0.4, that is inferred from IR emission or extinction measurements.

\subsection{The relation between Dust and ISM metallicity}

In our model, the fraction of metals locked up in dust follows that of the overall ISM metallicity
very closely, regardless of the star formation history of the ISM. This constancy between the dust
metallicity and that of the ISM is a result of the fact that the grain destruction
 timescale and the accretion timescale are both proportional to $\sigma_{ISM}$(r,t)/${\cal B}$(r,t) (see eq. 29 and
the discussion in \S 7).

Supporting evidence for our hypothesis that $\tau_{SNR}$/$ \tau_{accr}$ is only a weak function of time can be
found in the radial gradient of the dust$-$to$-$gas mass ratio of our Galaxy (see Figure 14a). Using {\it COBE}/DIRBE
observations of the large scale IR emission from the Galactic plane, Sodroski et al. (1997) constructed
a three$-$dimensional model of the IR emissivity of the Galaxy. In their model they derive the dust
abundance gradient, and find that it is equivalent, within the observational uncertainties, to the
metallicity gradient in the galactic disk. Our model results confirm these results. They show 
that the dust and ISM metallicity are tightly correlated.
The radial gradient in Z$_{dust}$ is identical to that of Z$_{ISM}$, maintaining a ratio of 0.4
at all Galacit radii. Since each annulus in the Galaxy has a distinct star
formation history, any time dependence in the
$\tau_{SNR}$/$\tau_{accr}$ ratio would result in a significant radial dispersion in the
Z$_{dust}$/Z$_{ISM}$ ratio, contrary to the results inferred from the infrared observations. 
 
External galaxies can give additional insight into the dependence of Z$_{dust}$ with Z$_{ISM}$.
In a recent paper, Lisenfeld
\& Ferrara (1997)  examined the relation between dust mass and gas metallicity in sample of dwarf
irregular and blue compact dwarf galaxies. The gas metallicity was determined from the O/H ratio in these objects, and
dust masses were derived from their 60 and 100 $\mu$m emission observed by the {\it IRAS} satellite. The sample presumably
represents these galaxies at various evolutionary stages, as manifested in their range of metallicities. 
In their analysis, Lisenfeld \& Ferrara (1997) found that (O/H) $\propto$ (M$_{dust}/M_{HI})^{0.54}$. For comparison, we plotted
our derived values of (O/H)$_{gas}$ versus Z$_{dust}$. The results are shown in Figure 14b, together with the data that was
presented in Figure 5 of Lisenfeld \& Ferrara (1997). The curves shown in the figure represent the trend in three different
regions of the Galaxy: the Galactic center, the solar circle, and the outer region of the Galaxy at R = 15 kpc. 
The results
for the various regions are essentially identical, giving a relation of (O/H) $\propto (M_{dust}/M_{ISM})^{0.77}$. The trend is
very similar to the data, but shows a systematic offset.
Various factors can contribute to this effect: (1) the dust masses in the galaxies are systematically lower than their
actual values. Dust masses were determined only from the 60 and 100
$\mu$m fluxes. The presence of emission from hot (equilibrium or stochastically heated) dust at 60
$\mu$m, can result in an overestimate of the dust temperature, and thus an underestimate of the dust mass. 
This effect can
introduce a factor of $\sim$ 2
$-$ 3 error in the actual dust mass. This underestimate can be systematic,
correlating with the abundance of hot dust, and perhaps with the star formation rate, and hence the metallicity of the system;
(2) the gas phase oxygen abundances in our model is systematically lower that observed in these galaxies. A significant
fraction of the oxygen depletion is caused by accretion in the ISM. The sample of dwarf galaxies is characterized by the
absence of dense molecular clouds, so accretion may not play an important role in the depletion of this element. The
success at which Lisenfeld \& Ferrara (1997) reproduced the observations {\it without} the inclusion of accretion in their
model, may lend some support to this explanation; and finally (3) the range of metallicities in these galaxies may not
represent a smooth evolutionary sequence, and actually reflect the stochastic nature of their chemical evolution. This effect
will introduce a spread in their metallicity, and consequently in their dust abundance as well.

\subsection{Evolution of the Dust Composition}    

Figure 15 depicts the evolution of the carbon and silicate dust mass surface density (Fig. 15a) and the evolution of 
the ratio of these two quantities (Fig 15b) as a function of time at the Galactic center (solid line), and in the
solar circle (dashed line). 
The evolution of the silicate and carbon dust abundances reflects the result of the
convolution of the stellar birthrates with
 the dust production yields. At all times the carbon-to-silicate dust mass ratio is less than unity since SNe, the main
interstellar dust sources, produce more silicate than carbon dust. The changes in the relative abundances of carbon and silicate
grains are the result of the delayed recycling of elements and dust back into the ISM by low mass stars. This delayed recycling
effect becomes  more evident
when the evolution of the carbon-to-silicate dust mass ratio is depicted as a function of time (Fig 15b). The stellar birthrate
at the Galactic center peaks before carbon stars turn off the main sequence. Consequently, the first generation of
dust particles are primarily silicates. The birthrate at the solar circle is still increasing when the first carbon stars
peel off the main sequence and start injecting carbon dust into the ISM. This results in a large increase in the
carbon-to-silicate dust mass ratio during the main sequence turnoff of carbon stars. The ratio
decreases after the last carbon$-$star progenitors leave the main$-$sequence.

These results have interesting implications for the possible classes of dust compositions in
external galaxies or young star forming regions evolving from an initially metal$-$free gas. According to the models
presented here, dust compositions, and hence extinction laws in external galaxies, should fall into three categories: 
(1) young galaxies or star forming regions should
 have overall low extinction characterized by a very weak 2175 \AA\ extinction bump (which is commonly
attributed to graphite grains). Such peculiar UV extinction has been observed with the {\it UIT} in the Small
Magellanic Clouds (Pr\'evot et al. 1984, Cornett et al. 1994). The lack of the 2175 \AA\ absorption
feature in the observed star forming field was attributed by these authors to the relatively young age
of the system, and the relative scarcity of evolved carbon stars; (2) intermediate age systems,
in which carbon stars are making a significant contribution to the ISM carbon abundance. The extinction law in these systems
should be characterized by the presence of a strong 2175 \AA\ extinction peak (compared to that of the Milky Way); and finally
(3) a phase where the extinction law should resemble that of the Milky Way galaxy.
The main assumption in this simple classification scheme is that the stellar mass spectrum is the same in all these
galaxies. However, the main underlying physical reasons that drive these differences are clear, and could be applied in the
modeling and analysis of any galaxy or star forming region.
  
%**************************************************
\section{SUMMARY AND ASTROPHYSICAL IMPLICATIONS}
%**************************************************

We have developed a model for the evolution of the composition and abundances of the elements and the dust
 in the Milky Way galaxy. The model consists of three components: (1) a dynamical infall model 
for the evolution of the stars and gas in the disk; (2) a chemical model for the evolution of the elements; and (3)
 a model for the evolution of dust in which dust is formed in quiescent stellar ejecta, and Type I and Type II SN ejecta,
and is destroyed by expanding SN blast waves. Accretion inside molecular cloud allows for grain growth in the ISM.
Table 1 summarizes the various model input parameters.

The dynamical infall model starts from a bulge and disk which grow by mass infall from zero mass, to the currently observed
bulge and disk mass surface densities.

The chemical evolution model reproduces the traditional observational tests summarized in Figure 4:
 the relative and absolute elemental abundances at the formation time of the sun, the age$-$metallicity relation, the
G-dwarf problem, and the Galactic [Fe/H] abundance gradient. Our model actually
reproduced the latter test better than models that do not include dust, since we calculate the actually observed
quantity, the {\it gas} phase abundance of Fe in the ISM, as a function of Galactic radius. Table 2 presents various model
output parameters.
 
We calculated the dust yield from low mass stars, and Type II and Type Ia SNe. Silicates are primarily produced in high
mass stars that become Type II supernovae, whereas carbon dust is primarily produced by stars in the ~ 2 - 5 \msun mass range
(Figures 6 and 7). The exact mass range of carbon stars depends on the initial stellar metallicity. 
Type Ia SNe produce iron in
excess of any other heavy element. Condensation in Type Ia SNe will therefore result in the injection of a significant amount
of Fe dust in a pure metallic form. The time dependence of the dust production rates of the various sources is given in
Figure 9, and the current values are presented in Table 3. 

Dust is destroyed by SNRs in the ISM, and we adopted the nominal grain destruction lifetimes calculated by Jones et al.
(1996). Grain destruction rates are higher than the rate at which they are injected into the ISM by the various sources.
Grain growth by accretion in molecular clouds plays therefore an important role in determining the dust mass, and we
adopted an accretion rate that is two times faster than the grain destruction rate. Grain lifetimes depend on the ISM mass and
the galactic SN rate, and are therefore evolving with time (Figure 8). We assumed that the accretion timescale has an
identical time dependence, preserving the ratio of the grain destruction and accretion timescale as a function of time.

Type II SNe constitute the most important source of interstellar dust.
SNR are the main source of grain destruction in the ISM, capable of destroying more dust than they produce. SN are
therefore  net destroyers of dust at the current epoch. However, early generations of supernovae are net producers
of interstellar  dust, since they expand in an initially dust$-$free or dust$-$poor interstellar medium.

We examined the origin of the elemental depletion pattern, to see if it reflects
condensation in the sources, accretion in clouds, or destruction in the ISM. In principle, if interstellar processing can be
neglected, condensation in stellar sources could explain the magnitude of the observed depletion pattern. However, even in this
case, it would be too simplistic to expect the depletion pattern to correlate with the condensation temperature. When the idea
was originally proposed by Field (1974), dust formation in Type Ia and Type II SNe was not an issue. Now that we know that these
objects synthesize the bulk of the silicates and iron injected into the ISM, other effects, such as explosion energies, the
amount of radioactivities, the degree of clumpiness in the ejecta, and dilution with gaseous ejecta from other sources can play
major roles in determining the depletion pattern. All this is actually a moot point, since the exchange of material between the
dust and gas phases of the ISM is so large, that all memory of the depletion pattern set up in the sources is erased in the
ISM. The depletion pattern therefore reflects the efficiency at which the various elements stick to and remain on the grains. In
the absence of any accretion in the ISM, the calculated elemental  depletions should reflect the elements  locked up in
refractory  grain cores. We examined the dependence of core depletions on grain destruction lifetimes, and found that even if
these lifetimes are increase by a factor of four,  the calculated core  depletions will still be significantly smaller than the
observed values.  This suggests either that grain destruction lifetimes are even longer than adopted in this paper, or that a
significant amount of the accreted mantle is reprocessed in the ISM into more refractory grain material. 

The silicate and carbon dust production rates depends strongly on stellar mass, with carbon dust produced primarily in low
mass carbon stars. The delayed recycling of the
gas and dust by these stars back to the ISM has important consequences for the evolution of the dust composition in various
regions of the Galaxy. The evolution of the ISM metallicity, and that of the carbon and silicate dust is presented in Figs. 12
(Galactic center) and Figure 13 (the solar circle). 
Overall, the evolution of the dust follows that of the metallicity. This is a simple consequence of the
fact that the relative importance of the grain destruction and grain accretion timescale remains constant in time. We
find that over most of the lifetime of the Galaxy, the mass of condensible elements locked up in dust is about 40\% of the total
mass of heavy elements in the ISM.
However, the relative abundance of carbon-to-silicate dust depends on the birthrate history of the
gas, and evolves as a function of time (Figure 15). From our models we can identify three epochs in the evolution of the dust
composition: during the first epoch, the dust population is silicate rich, since carbon stars have not yet evolved off the main
sequence; during the second epoch these stars inject carbon rich dust into the ISM, and the carbon-to-silicate mass ratio
increases significantly; the last epoch starts roughly when the lowest mass carbon star evolved off the main sequence, and is
marked by an increase in the abundance of silicate dust. As the figure shows, not all regions of the Galaxy
will undergo these phases, since they depend on the birthrate history of a given parcel of gas. 

An important hypothesis made in this paper, is that the accretion lifetime has an identical
time dependency as the grain destruction lifetime. The relative efficiency of these two processes is therefore constant
over Galactic history, ensuring a tight correlation between the dust abundance and metallicity. This hypothesis can be
tested by observing the radial dependence of the correlation between the dust-to-gas mass ratio and the metallicity. Large
scale far-infrared observations of the  Galactic plane suggest that the dust abundance follows the metallicity gradient.
Since each annulus in the Galaxy has a distinct star formation history, it is hard to imagine such correlation if the
relative efficiencies of grain destruction and accretions were widely time-variant. A plot correlating the dust mass with the
oxygen abundance in a sample of dwarf galaxies shows a similarity in the trend of the model and the data, but also the
presence of a systematic offset. This offset could be the the result of the combined effects of an underestimate in the
observed dust mass in these galaxies, the lack of significant oxygen accretion in their ISM, and the stochastic nature of
star formation and chemical evolution in these galaxies. 

If this hypothesis is indeed correct, then it is interesting to examine the possible processes that are
responsible for keeping the relative efficiencies of these two processes in check. Observations of external galaxies may
offer a clue. We examined the grain destruction timescales in M82, and found them to be extremely short, compared to the
duration of the starburst. M82 is therefore not merely a starburts galaxy, but an extremely efficient dust 
destroyer as well.
With most of the metals back in the gas phase, the post starburst region can cool very efficiently, so that the epoch of grain
destruction will be rapidly followed by an epoch of accretion onto any surviving grains. Even in the extreme case, in which all
the dust in the starburst region is destroyed, the injection of newly synthesized dust by low mass star, or the admixing of dust
from surrounding regions not affected by the starburst into the dust depleted region, will provide new nucleation centers onto which refractory
gas phase elements can accrete. So the relative efficiency of the grain destruction and accretion processes may not be constant
at any given instant in time, but constant on timescales averaged over a few hundred million years.
    The discussion above illustrates also the importance of including dust in chemodynamical models for the evolution of
galaxies (e.g. Samlar, Hensler, \& Theis 1997). These models consider the detailed interplay between star forming
processes, metal production, stellar evolution, and the energetics and phase changes in the ISM. Interstellar depletions, and
grain destruction will have an important effect on the cooling and energy balance of the ISM as well.
 
The results of this paper have important consequences for calculating the extinction in external galaxies. Young galaxies, or
star forming regions that underwent a recent burst of star formation, will be silicate rich, and their extinction law may be
characterized by the absence of the 2175
\AA\ UV extinction bump that is widely attributed to the presence of graphite grains. The SMC may be such a galactic system. Our
models, however, predict that some galactic systems, observed at just the right redshift, should have an anomalous extinction
law (relative to the average galactic one) characterized by an increase in the 2175 \AA\ bump. Since graphite or carbon dust has
a significantly larger UV-optical extinction, this result may have important consequence for the UV and optical appearance of
these systems, and for the derivation of various intrinsic galactic properties that are affected by dust extinction.

Finally we would like to point out that the models presented here provide a framework for the self-consistent inclusion of
the effects of dust in models for the population synthesis of galaxies.

\acknowledgments
During the course of this work I had the pleasure of
stimulating conversations with Cristina Chiappini, Matt Greenhouse,Ute Lisenfeld,  Harvey Moseley, Ant Jones, 
John Scalo, Randy Smith, Allen
Sweigart, and Frank Timmes. Special thanks are due to Frank Varosi who provided helpful programing advice, and to Ute
Lisenfeld for communicating her data on short notice. This research was supported by NASA Astrophysical Theory Program.

%\clearpage

%***************************** TABLES ******************************

%--------- Table 1 ---------------
\begin{deluxetable}{lll}
\tablecaption{Input Parameters for the Chemical Evolution of the ISM}
\tablehead{
\colhead{Parameter} & \colhead{Description} & \colhead{Units/Comments}  \nl 
  }
\startdata
%\underline{Evolution of ISM} & & \nl
$\sigma_{tot}$(r,t) & total surface mass density of the disk (stars+ISM) & M$_{\odot}$ pc$^{-2}$   \nl
$\sigma_{ISM}$(r,t) &  surface mass density of the ISM (gas+dust) & M$_{\odot}$ pc$^{-2}$   \nl
$\sigma_{ISM}$(A,r,t) &  surface mass density of element $A$ in the ISM & M$_{\odot}$ pc$^{-2}$   \nl
${\cal B}$(r,t) & star formation rate &  M$_{\odot}$ pc$^{-2}$ Gyr$^{-1}$   \nl
M$_{ej}(A,M)$ & mass of element $A$ ejected by star of mass $M$ & M$_{\odot}$   \nl
t$_G$ & age of the Galaxy & 12 Gyr   \nl
k, n & powers of the relation ${\cal B}\ \sim \nu \sigma_{tot}^k \sigma_{gas}^n$ & 0.5,\ 1.5   \nl
$\nu$  &  efficiency of the star formation rate & 1 Gyr$^{-1}$  \nl
R$_{SNIa}$ & the Type Ia SN rate &  pc$^{-2}$ Gyr$^{-1}$  \nl
R$_{SNII}$ & the Type II SN rate &  pc$^{-2}$ Gyr$^{-1}$  \nl
R$_{G}$ & total radius of the Galaxy & 15 kpc \nl
R$_{\odot}$ & solar distance to Galactic center & 8.5 kpc \nl
$\sigma_0$ & total surface mass density at Galactic center & 2.7 10$^3$ M$_{\odot}$ pc$^{-2}$ \nl
R$_{bulge}$ & bulge scalelength & 1.30 kpc \nl
$\tau_{inf}$(bulge) & infall timescale at R $\leq$ 2 kpc & 1 Gyr \nl
$\sigma_{\odot}$ & total surface mass density at solar circle & 60 M$_{\odot}$ pc$^{-2}$ \nl
R$_{disk}$ & disk exponential scalelength & 3.5 kpc \nl
$\tau_{inf}(R_{\odot})$ & infall timescale at R = R$_{\odot}$ & 4 Gyr \nl
M$_G$ & the total mass of the Galaxy (bulge + disk) & 6.5\ 10$^{10}$ \msun \nl
$\phi$(M) & the SMS normalized so that  $\int_{M_l}^{M_u}\phi(M)\ dM =1$   \nl
M$_l$, M$_u$ &  mass limits of the SMS & 0.1 M$_{\odot}$, 40 M$_{\odot}$   \nl
M$_{b1}$, M$_{b2}$ & mass limits for binary systems & 3 M$_{\odot}$, 16 M$_{\odot}$   \nl
M$_w$ & lower mass limit for Type II SNe & 8  M$_{\odot}$  \nl
% & & \nl
%\underline{Evolution of Dust} & & \nl
$\sigma_{dust}$(r,t)  &  surface mass density of dust & M$_{\odot}$ pc$^{-2}$   \nl
$\sigma_{dust}$(A,r,t) & surface mass density of element $A$ locked up in dust & M$_{\odot}$ pc$^{-2}$    \nl
$\delta_{cond}^w$,$\delta_{cond}^I$,  & the condensation efficiency of
                                        element $A$ in the ejecta  &  see \S5.2 \nl
$\delta_{cond}^{II}$, $\delta_{cond}^{WR}$ & of stellar winds (w), Type I SN (I), Type II SN (II), & for  \nl
  & and WR stars (WR)  & details \nl
$\tau_{SNR}(R_{\odot},t_G)$ & average destruction timescale in ISM &   0.4, 0.22, 0.3 Gyr for \nl
  &  &  carbonaceous, silicate, \nl
  &  &  and iron grains \nl
$\tau_{accr}(R_{\odot},t_G)$ & average accretion timescale in ISM  & $\tau_{SNR}$/2 \nl
\enddata
%\tablenotetext{a}{ }  
\end{deluxetable}

%--------- Table 2 ---------------
\begin{deluxetable}{lll}
\tablecaption{Output Parameters for the Chemical Evolution of the ISM}
\tablehead{
\colhead{Parameter} & \colhead{Calculated} & \colhead{Observed}  \nl 
  }
\startdata

${\cal B}$(R$_{\odot}$,t$_G$)\ M$_{\odot}$ pc$^{-2}$ Gyr$^{-1}$ & 3.8 & 2$-$10 \tablenotemark{a}  \nl
R$_{SNIa}$(R$_{\odot}$,t$_G$)    pc$^{-2}$ Gyr$^{-1}$            & 3.65 10$^{-3}$ & \nl
R$_{SNII}$(R$_{\odot}$,t$_G$)   pc$^{-2}$ Gyr$^{-1}$            & 2.44 10$^{-2}$ & \nl
R$_{Ia/II}$ = R$_{SNIa}$/R$_{SNII}$ & 0.15 & \nl
Global Type Ia rate (per century) & 0.22 yr$^{-1}$ & 0.3-0.4 \tablenotemark{b} \nl
Global Type II rate (per century) & 1.4 yr$^{-1}$ & 0.45-2.2 \tablenotemark{b} \nl
$\beta$ & 0.0048 & \nodata  \nl
$\sigma_{gas}(R_{\odot},t_G)$   M$_{\odot}$ pc$^{-2}$ & 8.8 & 5.7$-$7.0 \nl
$\sigma_{gas}(R_{\odot},t_G)/ \sigma_{stars}(R_{\odot},t_G)$ & 0.15 & \nl
$\left({d\sigma(t)\over dt}\right)_{infall}(R_{\odot},t_G)$  M$_{\odot}$ pc$^{-2}$ Gyr$^{-1}$ & 0.8 & $<$ 1.0 \nl
Z$_{dust}$(R$_{\odot}$,t$_G$) & 0.0087 & 0.0073 \tablenotemark{c}\nl
Z$_{crb}$(R$_{\odot}$,t$_{\odot}$) & 0.0017 & 0.0027 \tablenotemark{c} \nl
Z$_{sil}$(R$_{\odot}$,t$_{\odot}$) & 0.0070 & 0.0046 \tablenotemark{c} \nl

\enddata
\tablenotetext{a}{Rana (1991) }
\tablenotetext{b}{van den Bergh \& Tamman (1991).} 
\tablenotetext{c} {We compare here the calculated dust abundances at the time the solar system formed to the total abundances of
refractory elements that can be locked up in dust, using solar system abundances. By doing this we circumvent the
problems associated with the uncertainties in the determination of the current set of reference abundances.}  
\end{deluxetable}

%-------- Table 3 ----------------
% \begin{deluxetable}{llll}
%\tablecaption{Dust Spectra/color temperatures in the Average ISM and Selected Cirrus
%Clouds\tablenotemark{a}}
%\tablehead{
% & \multicolumn{3}{c}{$I(\lambda)/I(100\ \mu$m)\ /\ $T_{color}$ (K)} \nl
%\colhead{$\lambda(\mu$m)} & \colhead{average ISM\tablenotemark{b}} 
%  & \colhead{Cloud 1\tablenotemark{c}} & \colhead{Cloud 2\tablenotemark{d}} 
%  }

\begin{deluxetable}{lllll}
\tablecaption{Dust Production Rates in the Solar Neigborhood at t = t$_G$\tablenotemark{a}}
\tablehead{
\colhead{Stellar Source} & \multicolumn{2}{c}
      {This work} & \multicolumn{2}{c}{Jones \& Tielens (1994)} \nl
\colhead{ } & \colhead{carbon} & \colhead{silicate}  & \colhead{carbon} & \colhead{silicate} \nl  
  }
\startdata
\underline{Quiescent mass loss} &  &  &  & \nl
C-rich stars &  2.8 & \nodata & 2.1 & \nodata \nl
O-rich stars &  \nodata & 3.7 & \nodata & 3.2 \nl
Wolf-Rayet stars  & 0.02 & \nodata & 0.06 & \nodata \nl
 & & & & \nl
\underline{Explosive mass loss} & & & & \nl
novae & 3 10$^{-3}$ & $<$ 3 10$^{-3}$ & 0.3 & 0.03  \nl
Type II SNe  &  1.5 & 7.0 & 2 & 12 \nl
Type Ia SNe  &  0.09 & 3.5 & 0.3& 2 \nl

 & & & & \nl
\underline{ISM processes} & & \nl
accretion in clouds\tablenotemark{b} & 52 & 380 & \nodata & \nodata \nl
 & & & &  \nl \hline
 & & & & \nl
total grain formation rate & 56  & 390 & & \nl

\enddata
\tablenotetext{a}{Entries are in units of 10$^{-3}$ \msun pc$^{-2}$ Gyr$^{-1}$ }
\tablenotetext{b}{For $\tau_{accr}$ = 0.2 Gyr and 0.11 Gyr for carbonaceous and silicate grains, respectively}  
\end{deluxetable}

%-------- Table 4 ----------------
\begin{deluxetable}{lll}
\tablecaption{Dust Destruction Rates in the Solar Neigborhood\tablenotemark{a}}
\tablehead{
\colhead{Process } & \colhead{carbon} & \colhead{silicate} \nl 
  }
\startdata
Star formation  & 7.8 & 32 \nl
 & & \nl
Supernova shocks\tablenotemark{b}  & 51 & 390 \nl
  & &  \nl \hline
  & & \nl
total grain destruction rates & 59  & 420   \nl
\enddata
\tablenotetext{a}{Entries are in units of 10$^{-3}$ \msun pc$^{-2}$ Gyr$^{-1}$ } 
\tablenotetext{b}{For $\tau_{SNR}$ = 0.4 Gyr and 0.22 Gyr for carbonaceous and silicate grains, respectively}
 
\end{deluxetable}

%---------- Table 5 --------------
\begin{deluxetable}{lllll}
\tablecaption{Comparison of Calculated Elemental Depletions with Observations\tablenotemark {a}}
\tablehead{
\colhead{Element } & \colhead{core} & \colhead{observed}  & \colhead{core + mantle} & \colhead{observed} \nl 
  }
\startdata
C & 	0.04 & \nodata &	0.49	 & 0.30$-$0.60 \nl
O  & 0.007 &\nodata &	0.52 & 0.36$-$0.60	   \nl
Mg & 0.047 & 0.5$-$0.7&0.56 & 0.95 \nl
Si & 0.048 & 0.13$-$0.4&0.56  & 0.94$-$1.0  \nl
S & 0.048 & 0$-$0.1&0.56  & \nodata \nl
Fe & 0.07 & 0.65$-$0.78& 0.54& 1.0 \nl
\enddata
\tablenotetext{a}{Observed values are from Savage \& Sembach (1996). The range in the observations reflects the
range of values obtained by using solar or B-stars reference abundances} 
\end{deluxetable}

%***************************  FIGURES ************************

%---- 1
\clearpage
\begin{figure}
\caption{The stellar birthrate ${\cal B}$(r,t) as a function of time. Shown are the stellar
birthrates in the bulge, at R=0 (solid line), at the solar circle, R = 8.5 R$_{\odot}$ (dashed line), and the
disk-averaged birthrate (thick solid line).}
\end{figure}

%---- 2
\begin{figure}
\caption{The Galactic birthrate history as a function of Galactocentric radius. The thin solid line is the
stellar birthrate at t = 2 Gyr, and the thick solid line represents the birthrate at the current epoch, t = 12
Gyr. Curves intersecting the y-axis between these two curves represent the Galactic birthrates at
$\Delta$t = 2 Gyr intervals, increasing from the top to the bottom curve.}
\end{figure}

%---- 3
\begin{figure}
\caption{The evolution of the mass surface density in the bulge (at R= 0) and at the solar circle ( R = 8.5
R$_{\odot}$), as a function of time. The total mass density (stars + ISM; solid lines) reaches a current value determined by 
the observed stellar density of the bulge and disk components of the Milky Way. The density of the ISM (dashed line) 
peaks at an early epoch, determined by the infall rate, and declines slowly thereafter.}
\end{figure}

%---- 4 a-d
\begin{figure}
\caption{Standard observational tests for chemical evolution models: (a) Predicted ISM abundances of select elements at 
t = 7.45 Gyr, normalized to their solar value. The horizontal dotted lines indicate where the values of the normalized
abundances are a factor of two below and above solar value; (b) the evolution of the iron mass fraction, Z(Fe) as a function of
time. Model results (solid line) are compared with the observational constraints; (c) the relative frequency of stars of
 a given metallicity versus metallicity (the G-dwarf problem); (d) the radial gradient of metallicity compared to
observations. The inclusion of dust in the model actually improves the fit to the observations which only sample the
iron that is in the gas phase of the ISM.}
\end{figure}   

%---- 5
\begin{figure}
\caption{The Galactic supernova rates at he Galactic center and solar circle as a function of time. The Type II
supernova rate is shown as a solid line, and the Type Ia rate as a dashed line. The relative rates were normalized to a
value of SN Ia/SN II = 0.15 at the current epoch.}
\end{figure}

%---- 6
\begin{figure}
\caption{The carbon and oxygen yield of stars in the 1 $-$ 40 \msun mass range for: (a) stars with an initial
metallicity Z = 0.0040; and (b) stars with an initial metallicity Z = Z$_{\odot} = 0.20$. Also shown in the
figures is the C/O ratio in the stellar ejecta. For M $<$ 8
\msun, stars with C/O ratios $>$ 1 in their ejecta were assumed to produce carbon dust, whereas stars with a C/O ratio
$<$ 1, were assumed to produce silicate dust. Higher mass stars can produce both type dust particles, regardless of the
C/O ratio in their ejecta.}
\end{figure}   

%---- 7
\begin{figure}
\caption{The SMS-weighted yield of carbon and silicate dust as a function of initial stellar mass for: (a) stars with an initial
metallicity Z = 0.0040; and (b) stars with an initial metallicity Z = Z$_{\odot}$ = 0.020. The figure shows
that most of the carbon is produced by low mass stars in the $\approx\ 2 - 5$ \msun mass range, whereas silicate dust is
produced predominantly in stars with M $\gtrsim$ 10 \msun. }
\end{figure}

%---- 8
\begin{figure}
\caption{The evolution of $\tau_{SNR}$, the grain destruction timescale by SNRs, and $\tau_{accr}$, the timescale for
 grain growth by accretion, as a function of time. Both timescales have identical time dependence (see \S 6.2 and \S 7
in text) so their ratio is constant over Galactic history. The timescales are shown for the Galactic center and the
solar circle. }
\end{figure}   

%---- 9
\begin{figure}
\caption{ A comparison of the dust production rate by various sources: SN II (solid line), stars with M $<$ \msun (dotted
line), SN Ia (dashed line); to the destruction rate by astration (thick solid line). The various panels depict separately,
the silicate and carbon production rates at the Galactic center and solar circle. Values at the current epoch at the 
solar circle are listed
in Table 3. For sake of clarity, the grain destruction rate by astration was scaled down by a factor of 3.}
\end{figure}   

%---- 10
\begin{figure}
\caption{The fraction of the elemental abundance that is locked up in dust for three cases: case 1 (open squares) assumes 
that the dust abundance is not altered by interstellar processes. These depletions should reflect the effect of
condensation in the sources; case 2 (open diamonds) takes only grain destruction into account and ignores grain growth by
accretion. The depletion in this case should reflect core depletions; case 3 (filled circles) includes the effect of grain
destruction and accretion. These depletions should reflect core $+$ mantle depletions inthe ISM.}
\end{figure} 
 
%---- 11
\begin{figure}
\caption{Core depletions are calculated as a function of $\tau_{SNR}$/$\tau_o$, where $\tau_o$ is the nominal value
 given by Jones et al. (1996), and equal to 0.22 Gyr for silicates, 0.4 Gyr for carbon dust, and we adopted an 
intermediate value of 0.3 Gyr for pure iron dust particles. }
\end{figure}

%---- 12
\begin{figure}
\caption{The evolution of the ISM metallicity Z$_{ISM}$, and dust metallicity Z$_{dust}$ as a function of
time at the Galactic center. Also shown are the separate contrubutions of carbon and
silicate dust to Z$_{dust}$, and the ratio Z$_{dust}$/Z$_{ISM}$.}
\end{figure}   

%---- 13
\begin{figure}
\caption{Same as Figure 12 in the solar circle.}
\end{figure}   

%---- 14
\begin{figure}
\caption{(a) the carbon-to-silicate mass ratio (solid line), and Z$_{dust}$, the total dust$-$to$-$gas mass ratio
 (dashed line) at the present epoch, as a
function of Galactocentric radius. Carbon dust is less abundant at the Galactic center, but the effect is small. The
 dust metallicity decreases as a function of radius tightly maintaining the relation Z$_{dust}(R,t_G)=
 0.4 Z_{ISM}(R,t_G)$ as a function of radius R.  (b) model predictions for the dependence of the gas phase oxygen abundance
on dust content, are compared to observations in Blue Compact Dwarf galaxies (X) and Dwarf Irregulars (triangles).
Data were taken from Lisenfeld \& Ferrara (1997). The figure is discussed in \S 8.4 of the text.}
\end{figure}

%---- 15
\begin{figure}
\caption{ (a) The
evolution of silicate and carbon dust at the Galactic center (solid line) and in the solar neighborhood (dashed lines)
as a function of time.
Peak abundances of the two dust populations reflect the result of the convolution of the stellar birthrate and
 the dust production yields; (b) The carbon$-$to$-$silicate dust mass ratio as a function of time at the Galactic center 
 (solid line), and in the solar circle (dashed line). Dust abundance variations result from the combined effects 
 of the stellar birthrate history, and the delayed recycling of carbon dust by low mass stars into the ISM.
 More details can be found in \S 8.5 of the text}
\end{figure}    

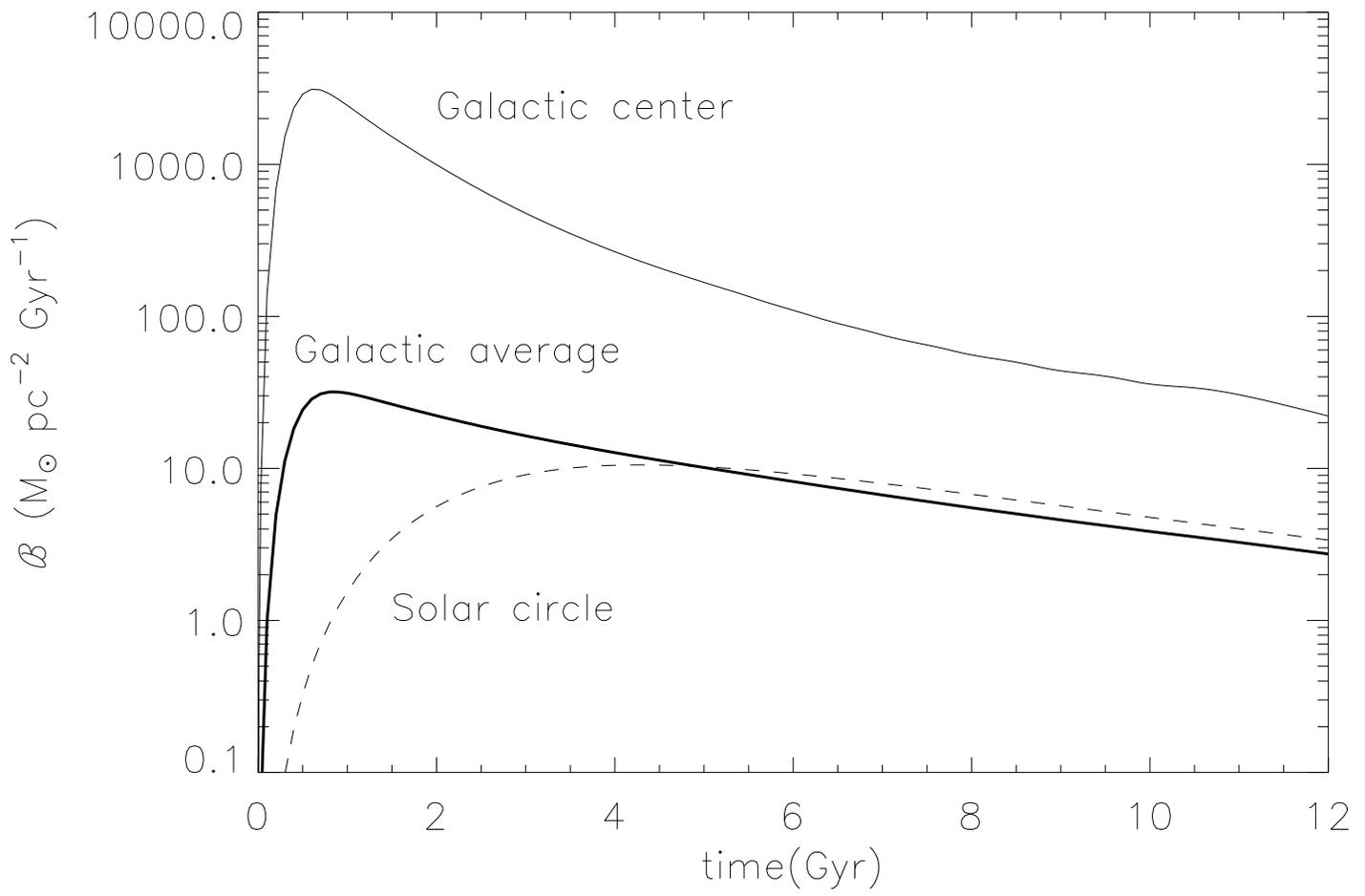
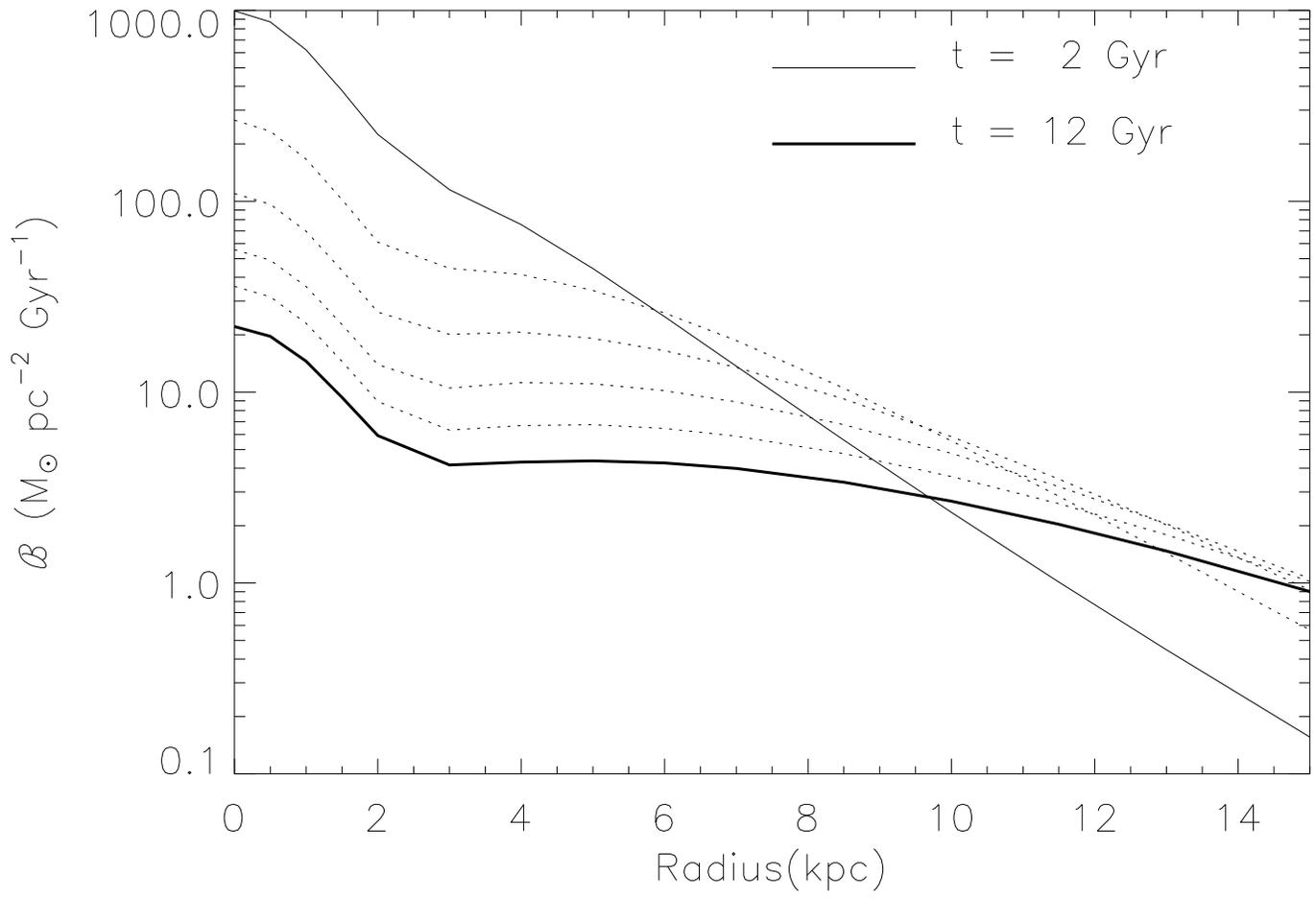
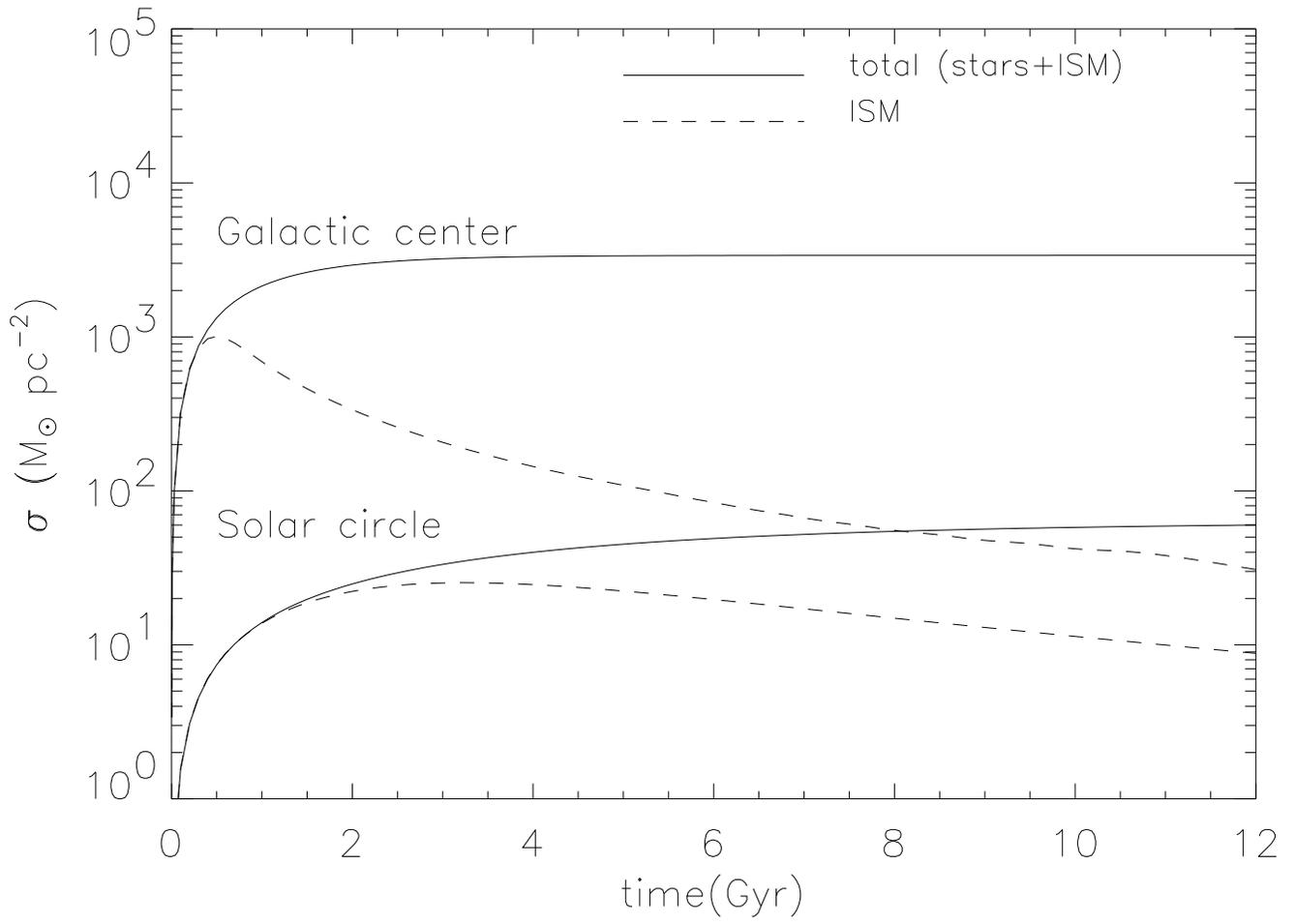
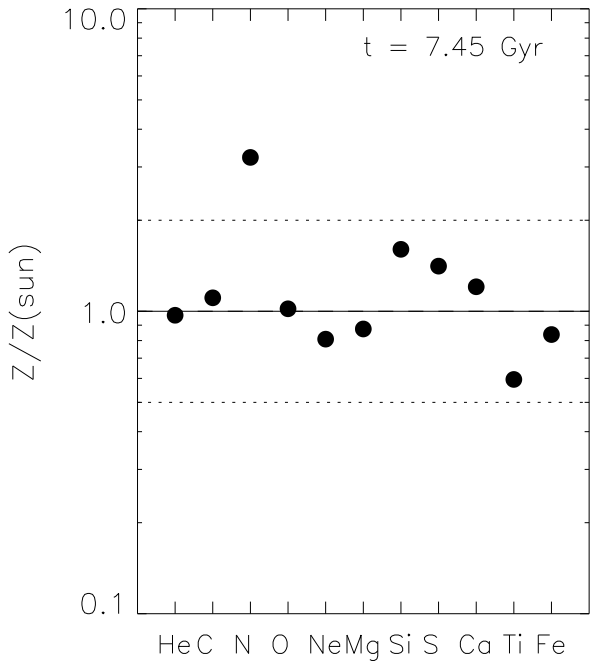
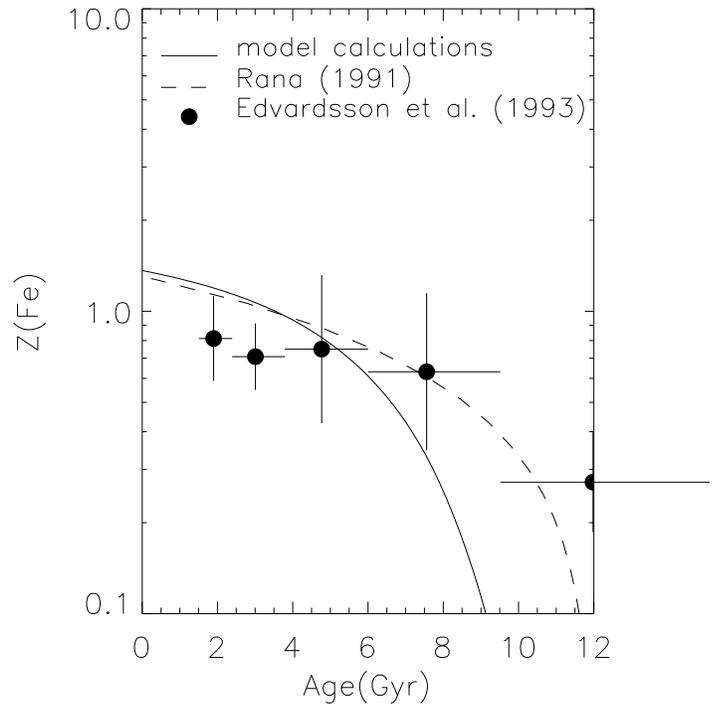
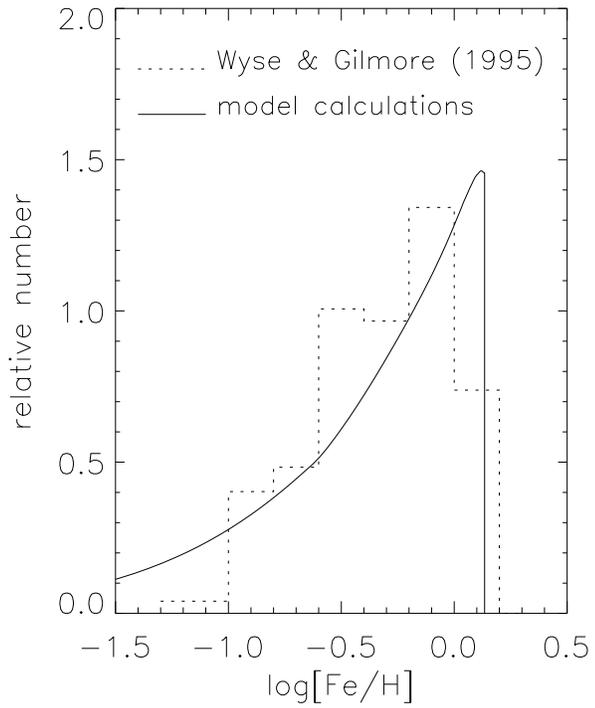
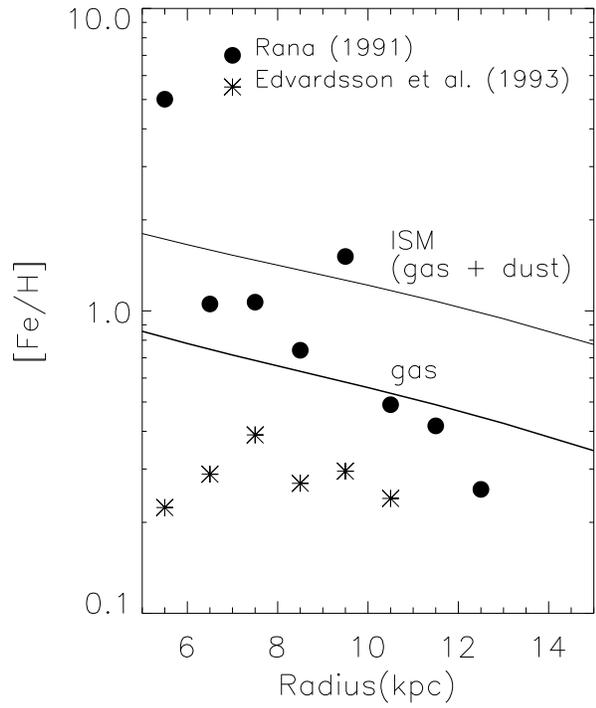
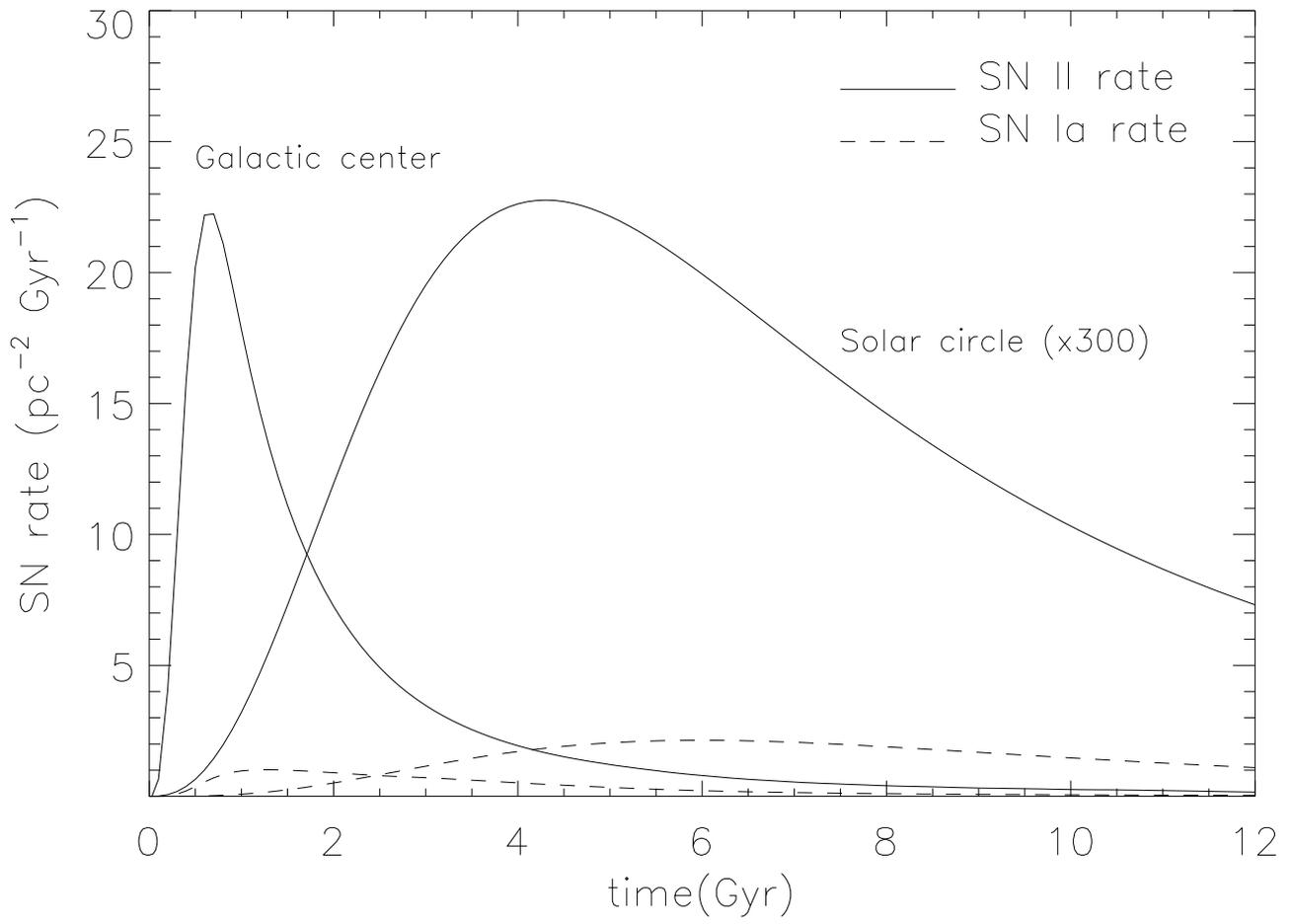
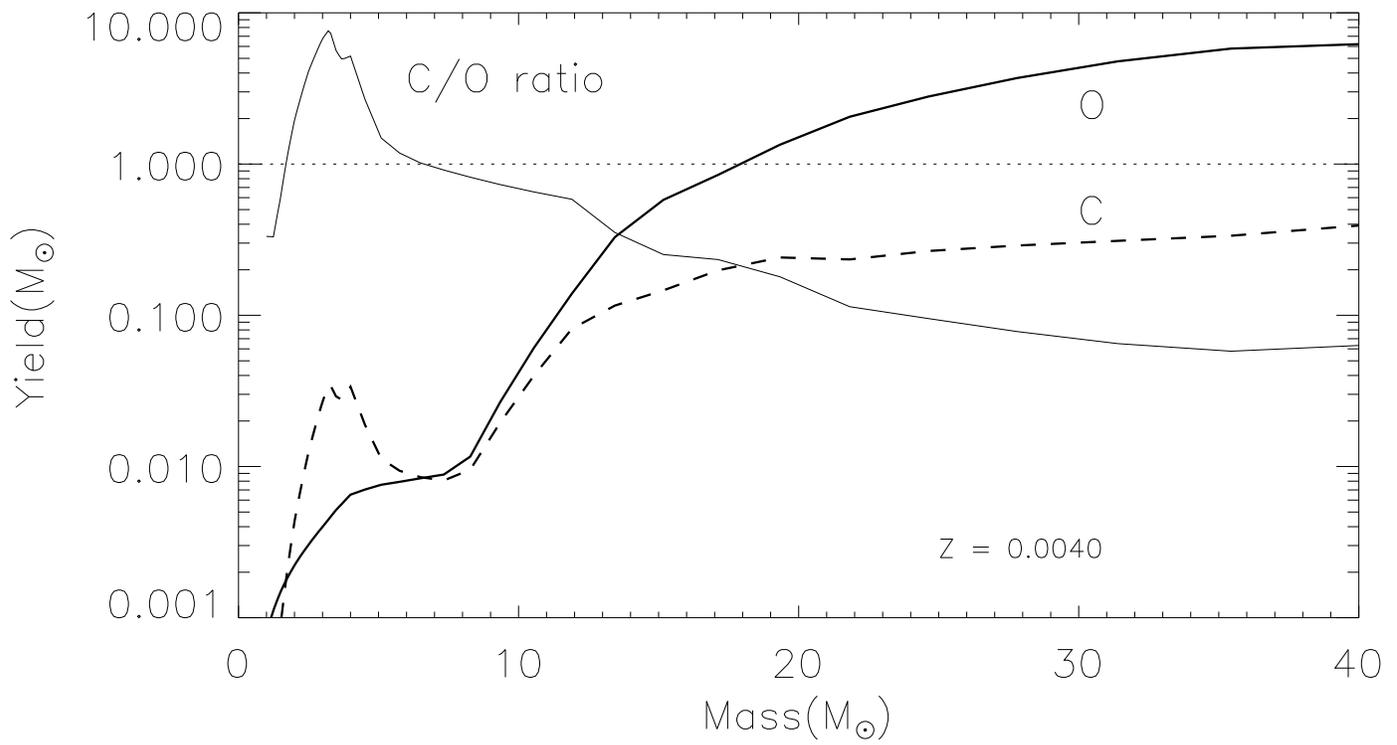
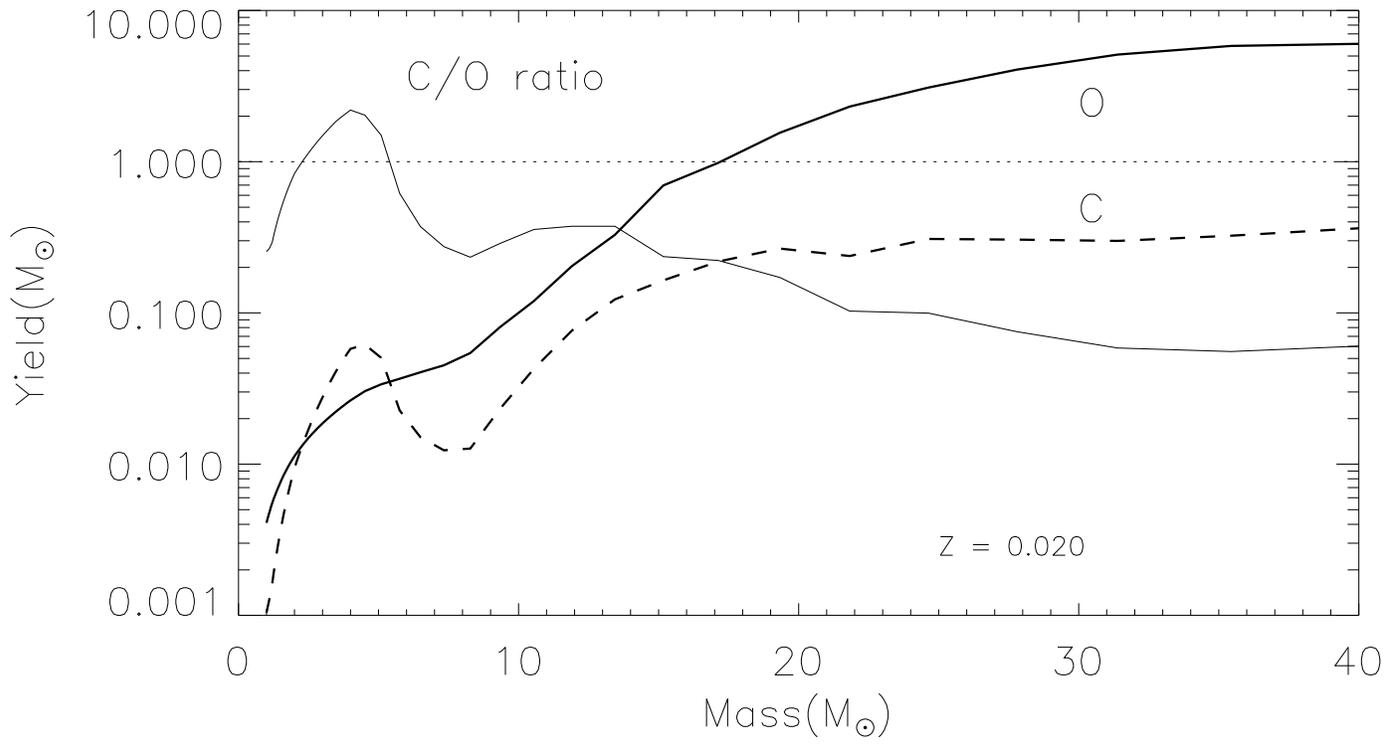
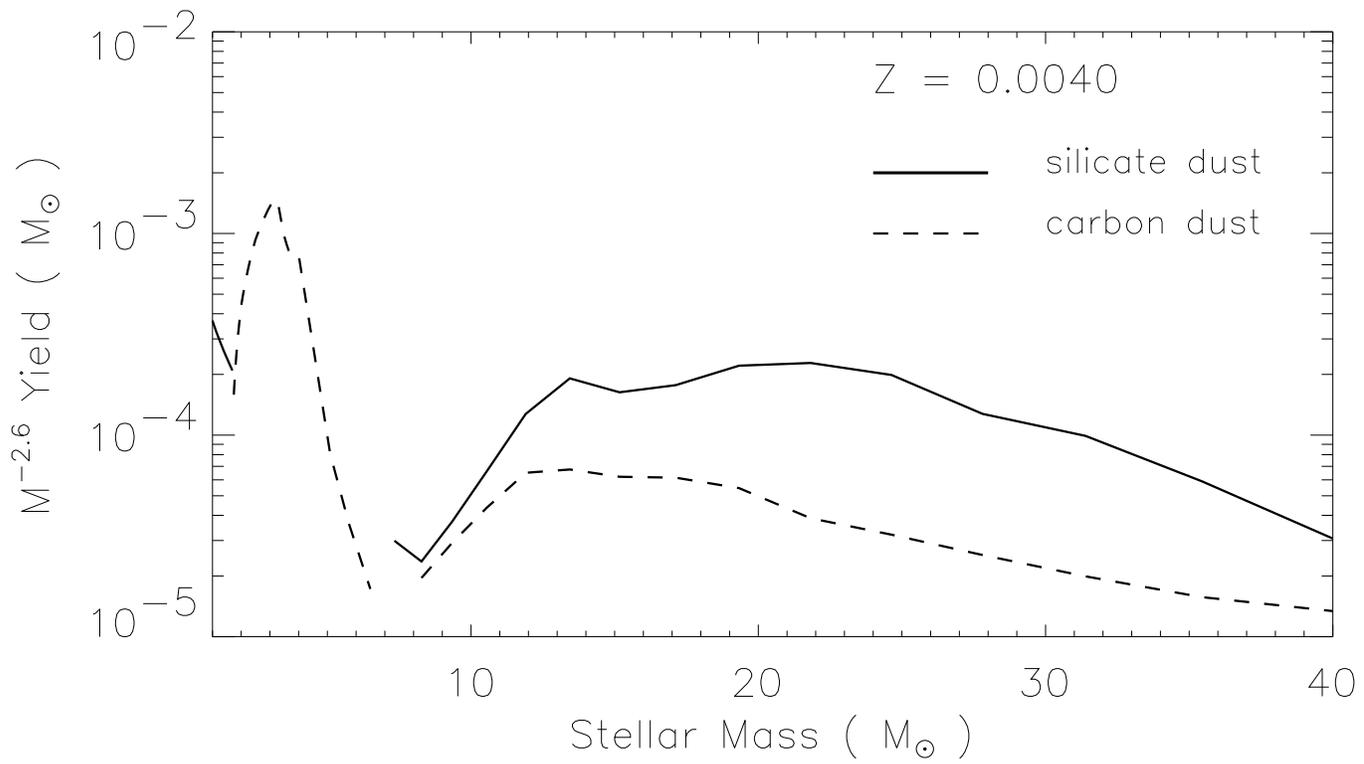
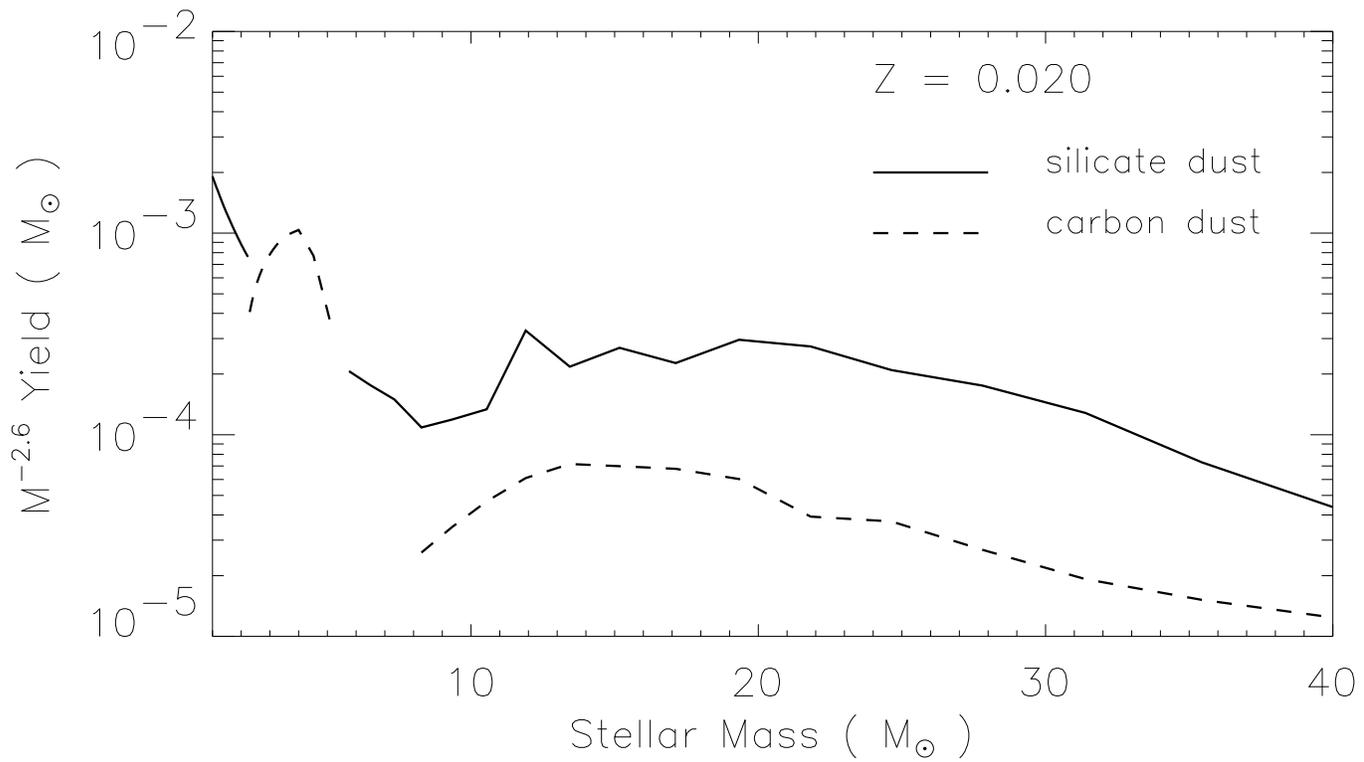
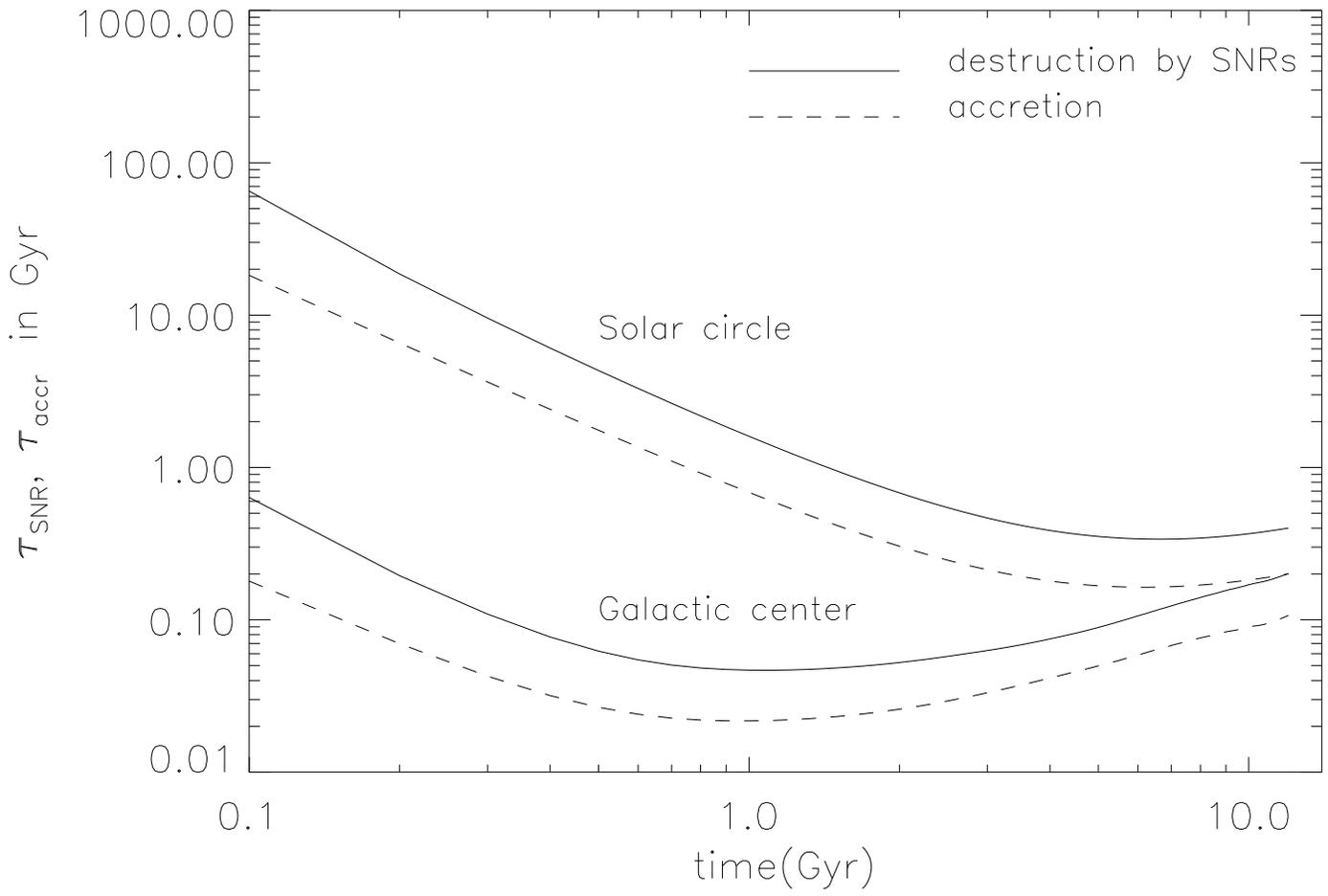
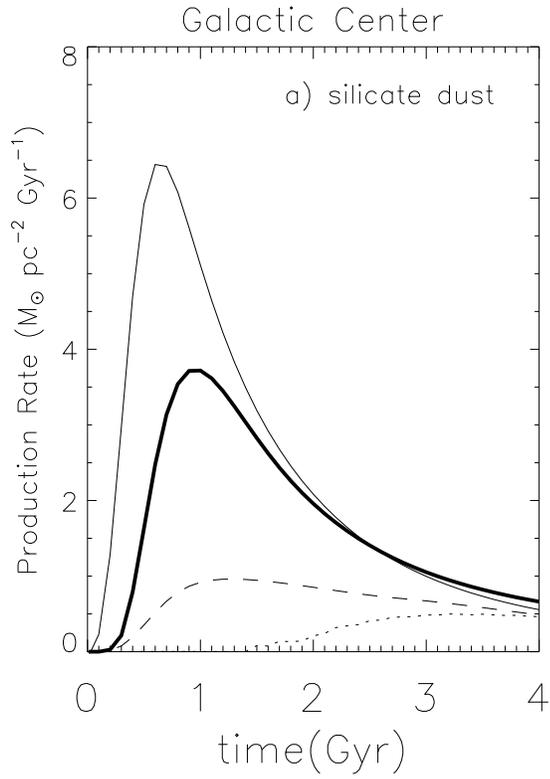
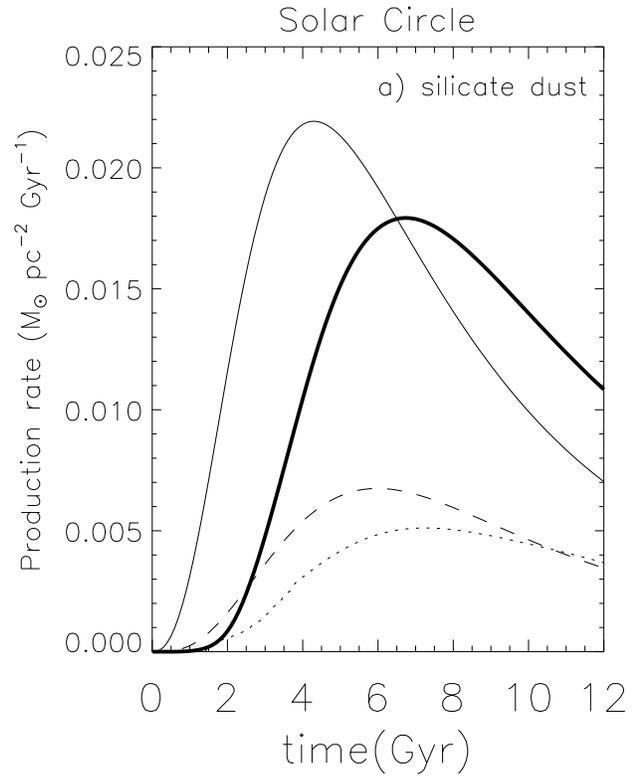
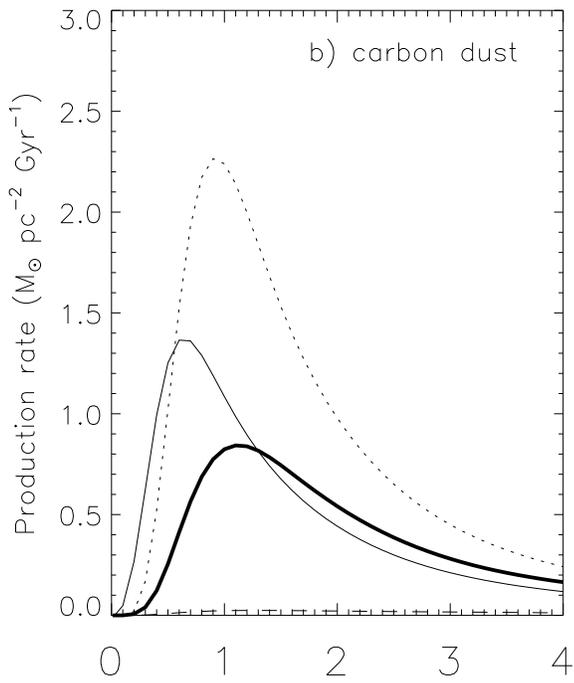
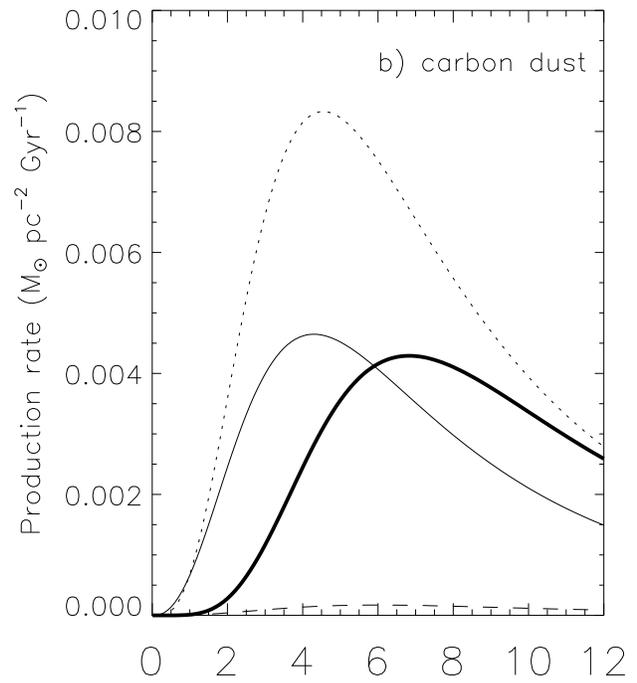
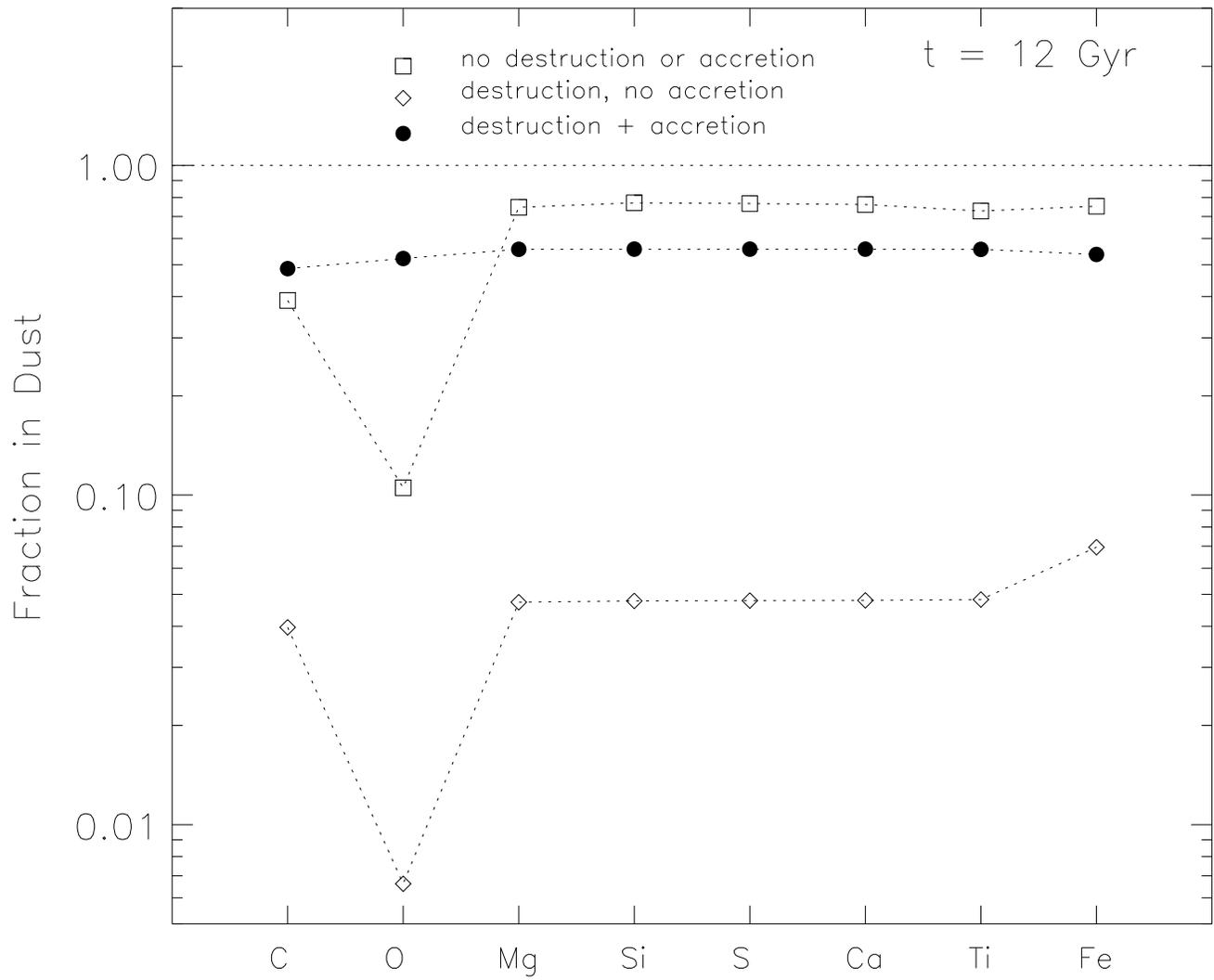
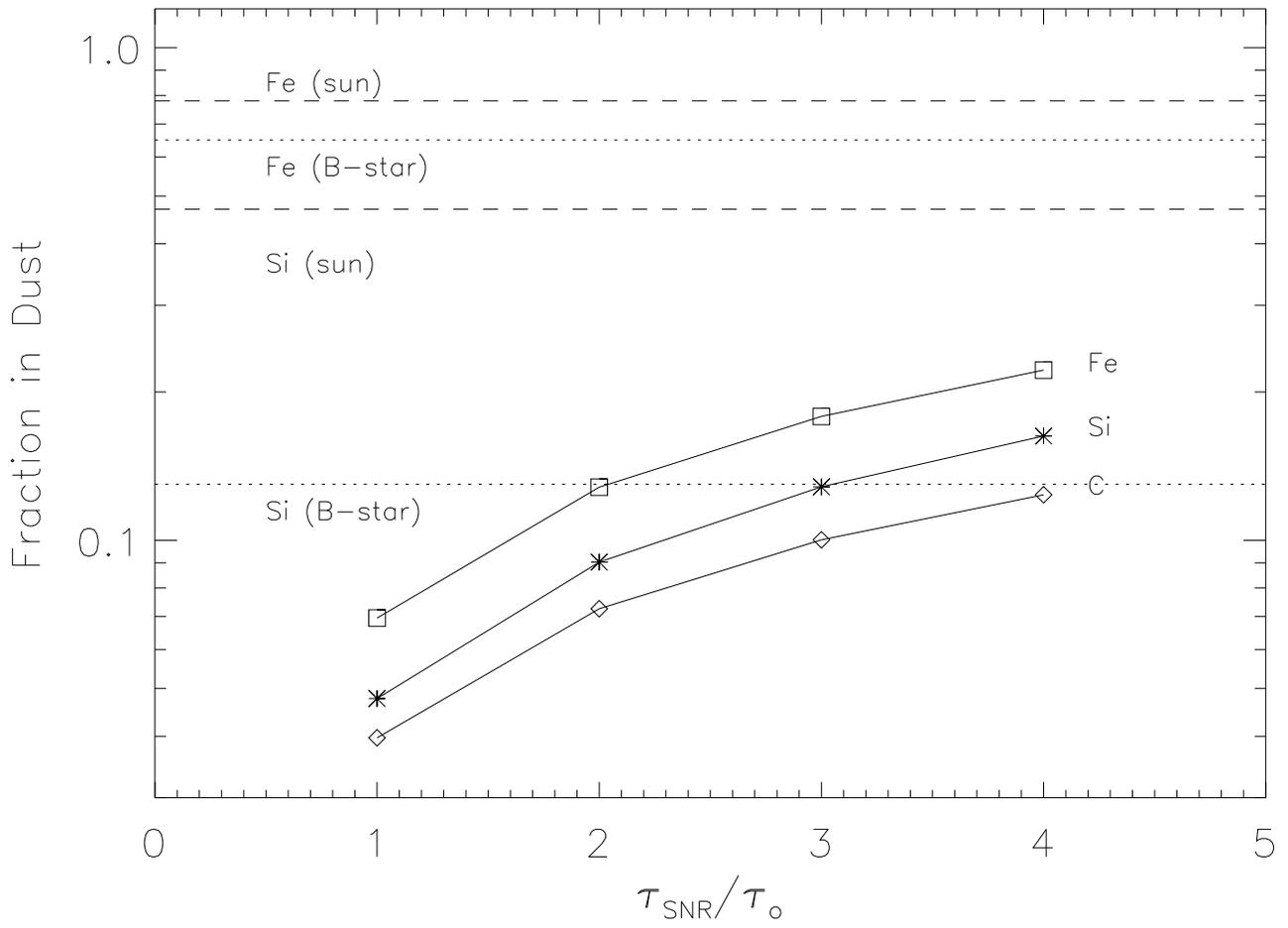
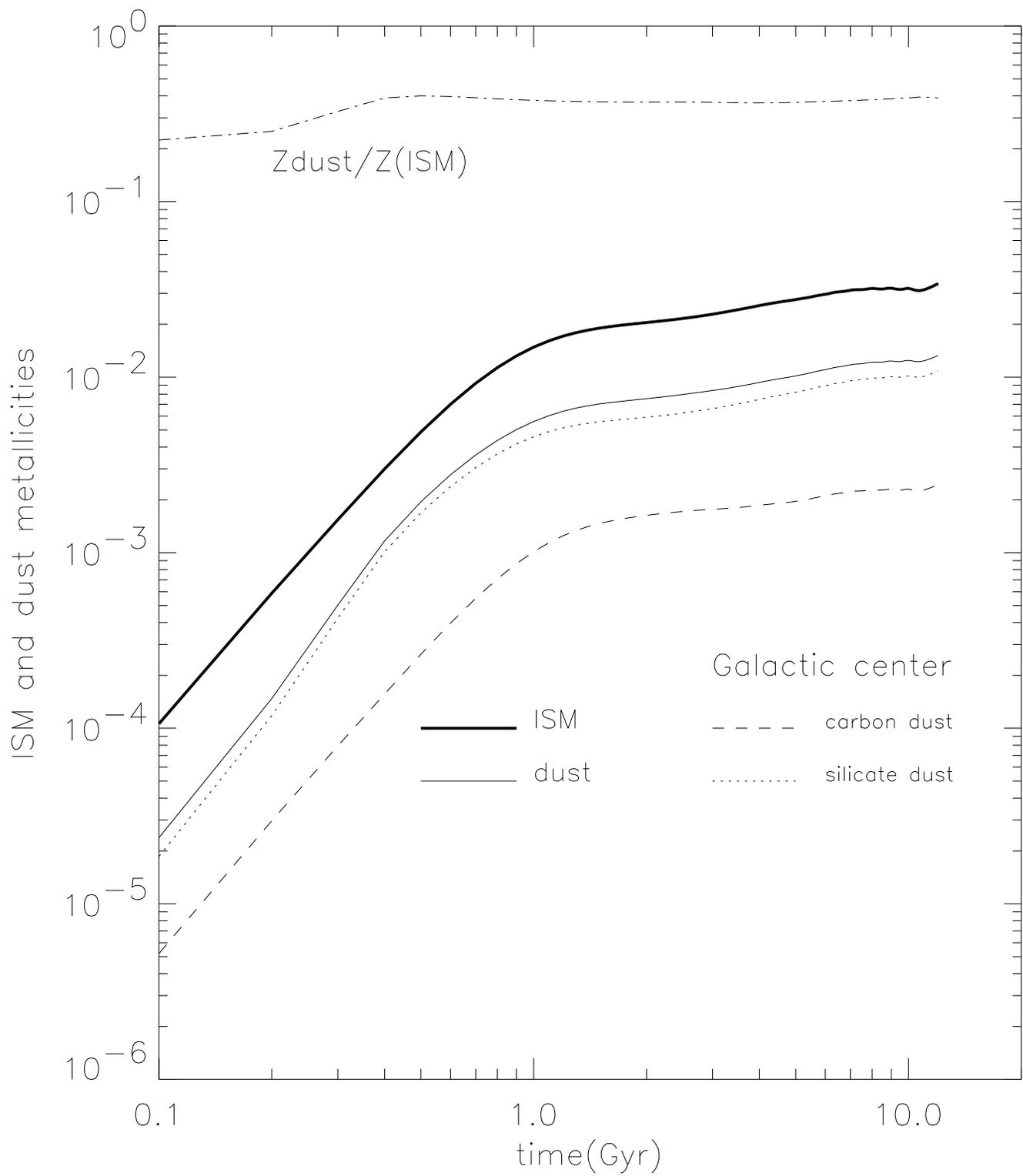
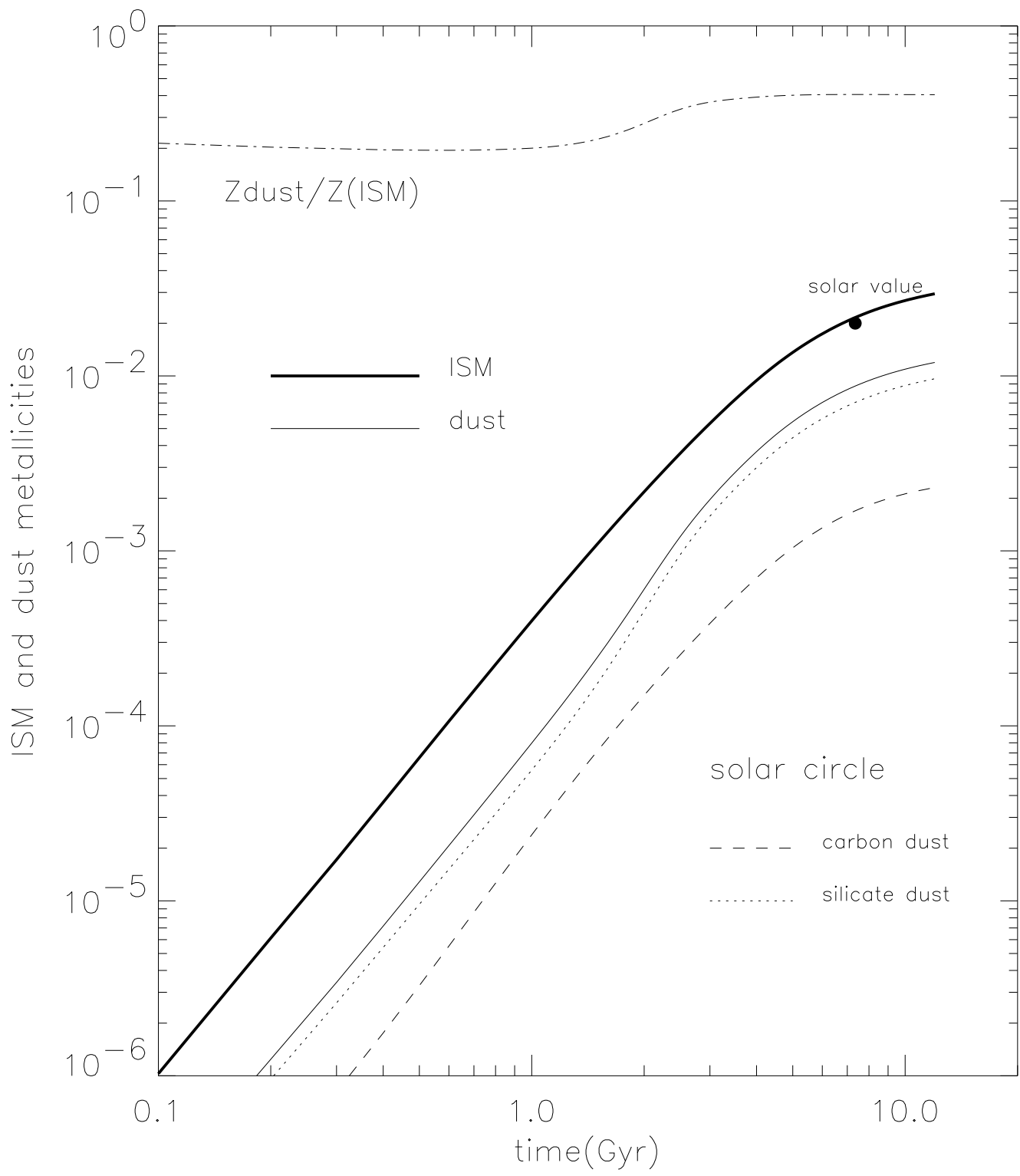
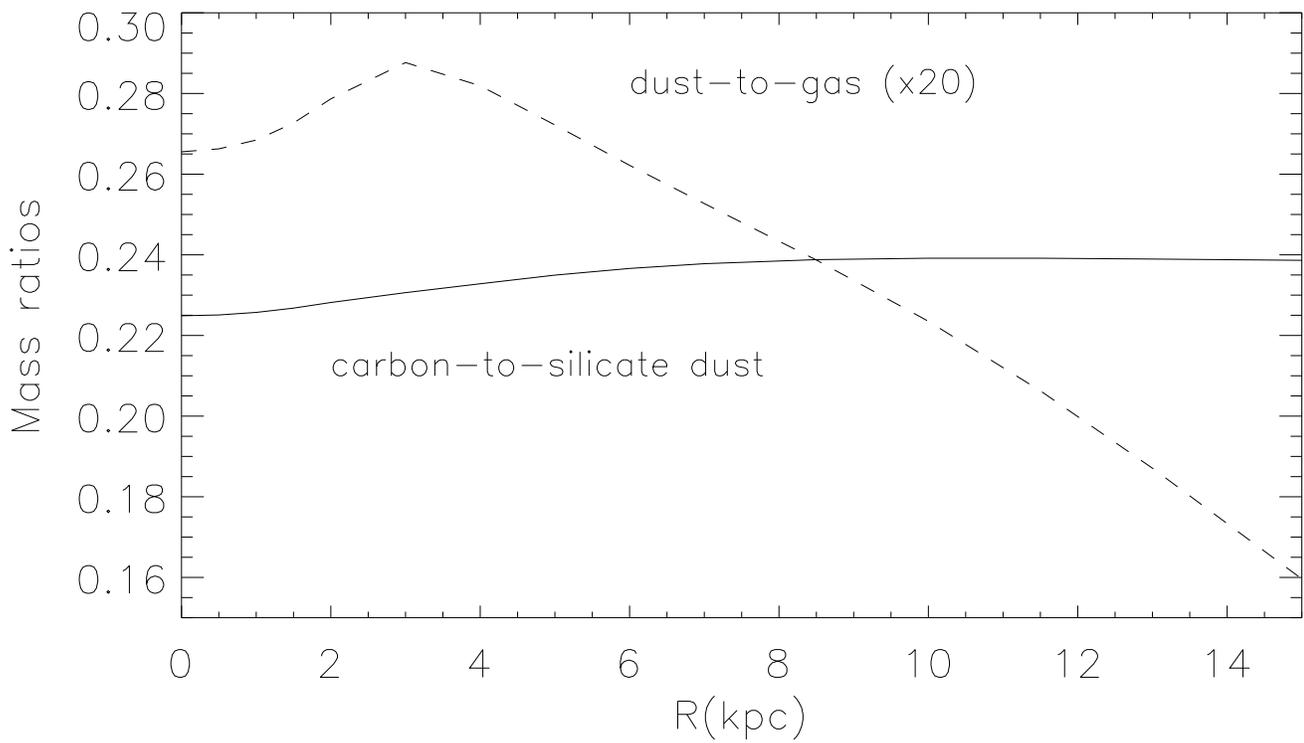
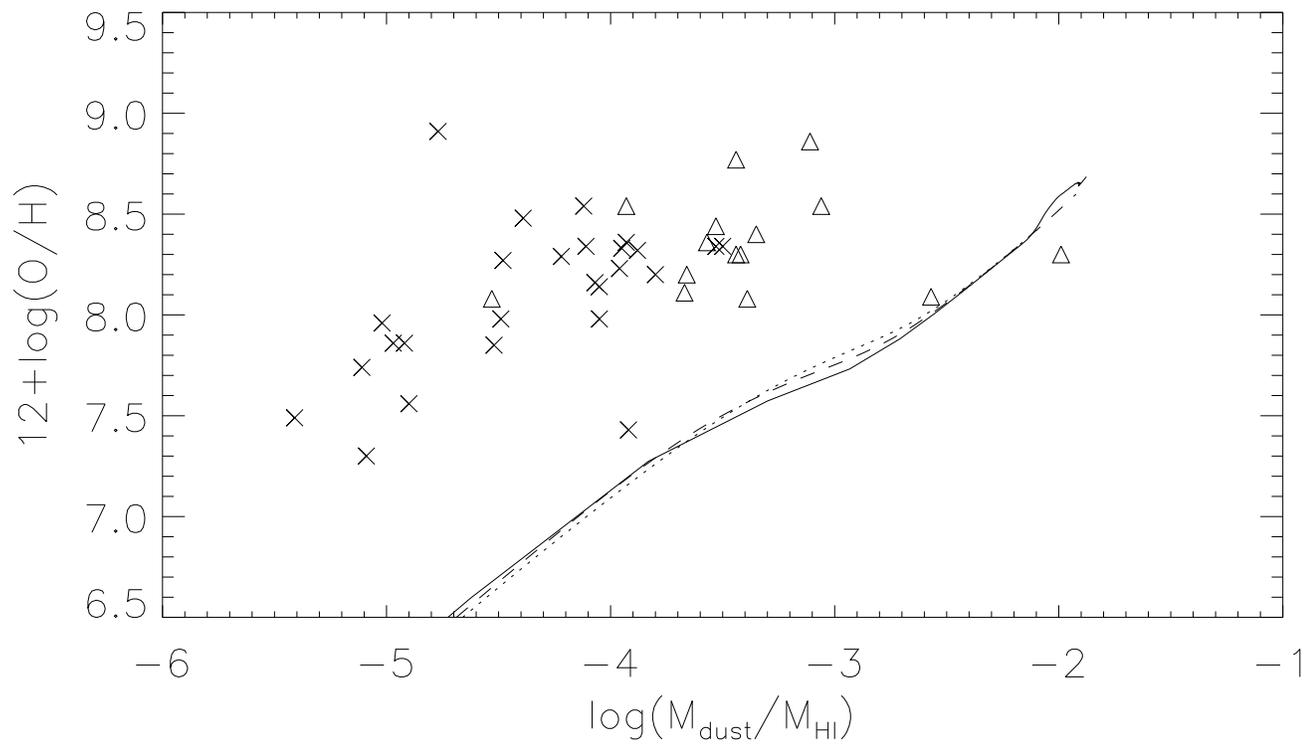
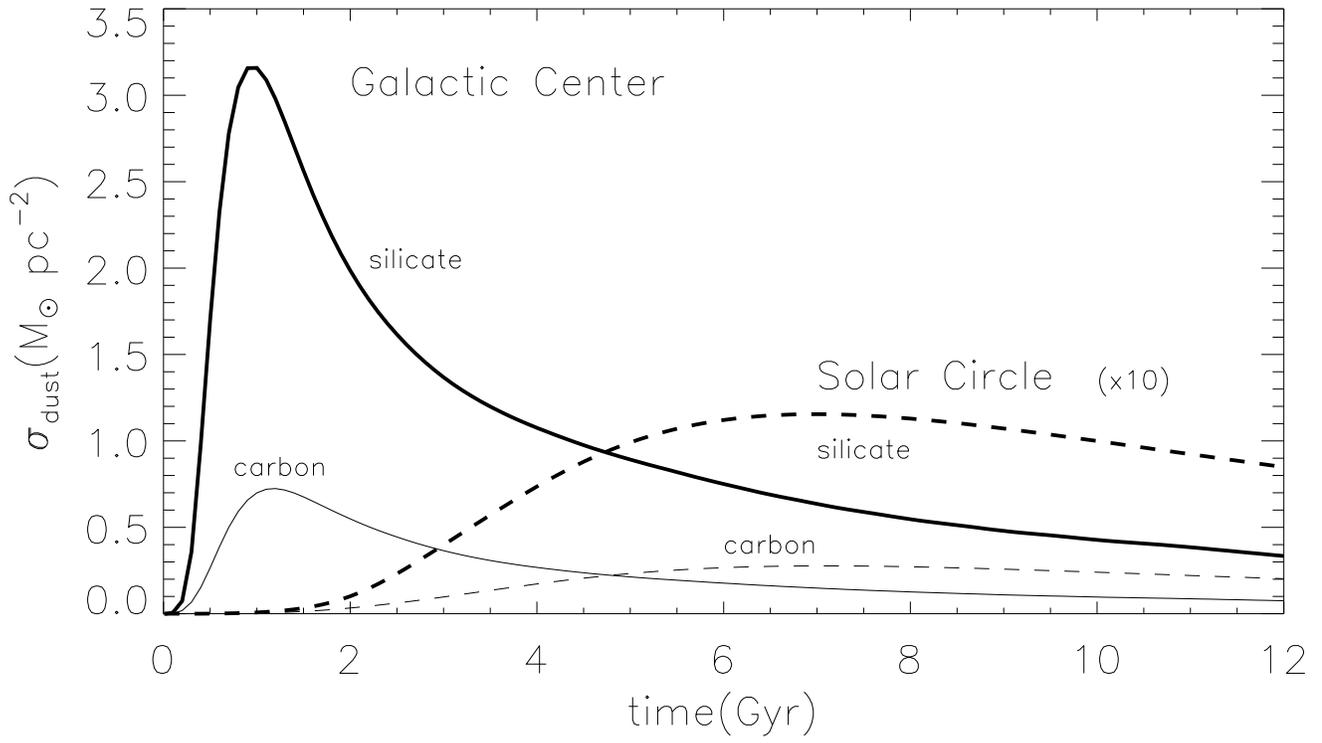
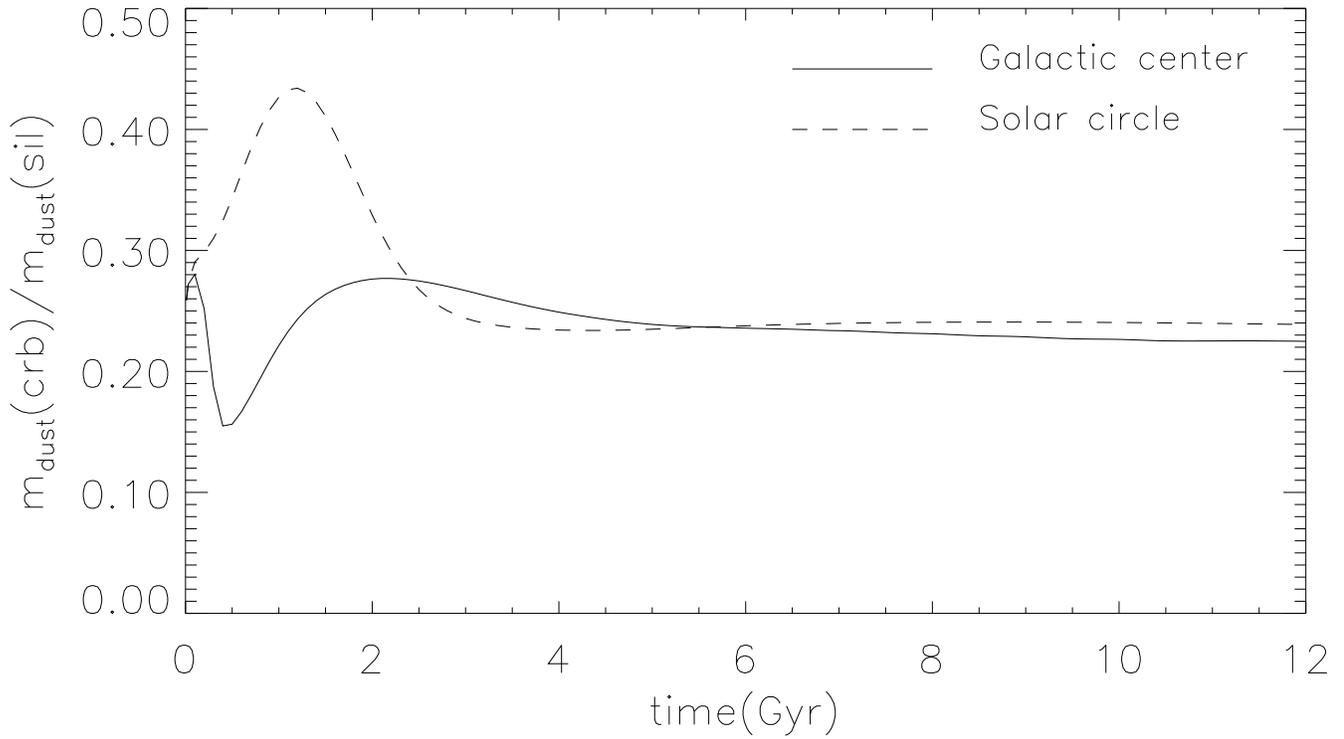
\end{document}